\documentclass[fleqn,usenatbib]{mnras}

\usepackage[T1]{fontenc}

\DeclareRobustCommand{\VAN}[3]{#2}
\let\VANthebibliography\thebibliography
\def\thebibliography{\DeclareRobustCommand{\VAN}[3]{##3}\VANthebibliography}

\usepackage{import}
\usepackage{graphicx}	
\usepackage{amsmath}	
\usepackage{bm}

\def\*#1{{\bm{#1}}}

\title[21-cm signal extraction with multiple antennas]{Sky-averaged 21-cm signal extraction using multiple antennas with an SVD framework: the REACH case}
\author[Saxena et al.]{Anchal Saxena,$^{1}$\thanks{E-mail: a.saxena@rug.nl}
P.\ Daniel Meerburg,$^{1}$
Eloy de Lera Acedo,$^{2,3}$
Will Handley$^{2,3}$
and \newauthor Léon V.E.\ Koopmans$^{4}$
\\
$^{1}$Van Swinderen Institute, University of Groningen, Nijenborgh 4, 9747 AG Groningen, The Netherlands\\
$^{2}$Cavendish Astrophysics, University of Cambridge, Cambridge, UK\\
$^{3}$Kavli Institute for Cosmology, Madingley Road, Cambridge CB3 0HA, UK\\
$^{4}$Kapteyn Astronomical Institute, University of Groningen, P.O Box 800, 9700 AV Groningen, The Netherlands
}

\date{Accepted XXX. Received YYY; in original form ZZZ}

\pubyear{2023}

\begin{document}
\label{firstpage}
\pagerange{\pageref{firstpage}--\pageref{lastpage}}
\maketitle

\begin{abstract}
In a sky-averaged 21-cm signal experiment, the uncertainty on the extracted signal depends mainly on the covariance between the foreground and 21-cm signal models. In this paper, we construct these models using the modes of variation obtained from the Singular Value Decomposition of a set of simulated foreground and 21-cm signals. We present a strategy to reduce this overlap between the 21-cm and foreground modes by simultaneously fitting the spectra from multiple different antennas, which can be used in combination with the method of utilizing the time dependence of foregrounds while fitting multiple drift scan spectra. To demonstrate this idea, we consider two different foreground models (i) a simple foreground model, where we assume a constant spectral index over the sky, and (ii) a more realistic foreground model, with a spatial variation of the spectral index. For the simple foreground model, with just a single antenna design, we are able to extract the signal with good accuracy if we simultaneously fit the data from multiple time slices. The 21-cm signal extraction is further improved when we simultaneously fit the data from different antennas as well. This improvement becomes more pronounced while using the more realistic mock observations generated from the detailed foreground model. We find that even if we fit multiple time slices, the recovered signal is biased and inaccurate for a single antenna. However, simultaneously fitting the data from different antennas reduces the bias and the uncertainty by a factor of 2-3 on the extracted 21-cm signal.
\end{abstract}




\section{Introduction}
\label{sec:intro}
The Cosmic Dawn (CD) and Epoch of Reionization (EoR) are one of the least known eras in the history of the Universe. This was the time when the first sources of light were formed, which emitted high energy X-ray and UV radiation, which in turn heated and reionized the intergalactic medium \citep[see e.g.][for reviews]{2001PhR...349..125B, 2006PhR...433..181F, Pritchard_2012}. Observations from the absorption spectra of high redshift quasars \citep{2001AJ....122.2850B, 2003AJ....125.1649F, boera19} and Thomson scattering optical depth from the Cosmic Microwave Background (CMB) \citep{2003ApJ...583...24K, 2011ApJS..192...18K} suggest that it was an extended process that took place between $z \sim 50 - 5$ \citep{Alvarez_2006, 2015ApJ...802L..19R, 2015ApJ...811..140B}. However, these indirect probes do not provide a very clear understanding of the physical processes occurring during the CD-EoR, such as the exact redshift range of primordial star formation, star formation efficiency, sources responsible for ionization, and the luminosity of first X-ray sources.

The 21-cm line associated with the spin flip transition in the ground state of the hydrogen atom is the most direct and promising approach to probe these eras. Motivated by the potential of the 21-cm line to understand the astrophysical processes of the infant universe, a large number of experiments are ongoing to detect and characterize this signal. These experiments can be classified into two different categories. The first are interferometric experiments, including those using the GMRT \citep{2013MNRAS.433..639P}, HERA \citep{DeBoer_2017}, LOFAR \citep{mertens20, 10.1093/mnras/staa487}, MWA \citep{barry19, li19}, PAPER \citep{kolopanis19} and SKA \citep{2015aska.confE...1K, mellema2015hi}, which are designed to measure the spatial brightness temperature fluctuations in the 21-cm signal using various Fourier statistics. Complementary, there are experiments to detect the sky-averaged 21-cm signal. These experiments include BIGHORNS \citep{Sokolowski_2015}, EDGES \citep{2018Natur.555...67B}, LEDA \citep{Price_2018}, PRIZM \citep{https://doi.org/10.48550/arxiv.1806.09531}, REACH \citep{2022NatAs...6..984D}, SARAS \citep{2018ApJ...858...54S, https://doi.org/10.48550/arxiv.2112.06778}. Such experiments are simpler to design than interferometric experiments and require shorter integration time, but the calibration is more challenging. The dynamic range in the foreground vs.\ 21-cm signal is much larger in global experiments, by several orders of magnitude. This paper concerns the latter approach for the Radio Experiment for the Analysis of Cosmic Hydrogen (REACH)\footnote{\url{https://www.astro.phy.cam.ac.uk/research/research-projects/reach}}.

Several years ago, the Experiment to Detect the Global EoR Signature (EDGES) reported a detection of an absorption trough centered at 78 MHz \citep{2018Natur.555...67B}. However, the shape and depth of the identified signal are quite different from the expectations of standard cosmological and astrophysical models. This unexpected feature either calls for an exotic explanation that could be achieved by enhancing the contrast between the radio background temperature and the spin temperature of the 21-cm signal or is still some unknown systematic effect in the data. Two physically motivated scenarios for the former explanation include the over-cooling of hydrogen gas due to interactions with dark matter \citep{barkana18, Berlin, Liu19, barkana18a} and the presence of an enhanced radio background other than the CMB \citep{ewall18,feng18,fialkov19,ewall20,reis20}. However, a re-analysis of the EDGES data has shown that unaccounted systematics can distort or mask the signal \citep{2018Natur.564E..32H, 2019ApJ...880...26S, 2020MNRAS.492...22S, Bevins2020}. There could also be inadequacies in the foreground model used to fit the observations, which could account for the observed trough \citep{Tauscher_2020a}. The measurement of the spectrum of the radio sky in 55-85 MHz band reported by \citet{https://doi.org/10.48550/arxiv.2112.06778} shows, albeit at $\sim2\sigma$ confidence level, that the detection of the signal claimed by EDGES might not be entirely of astrophysical origin.

In order to make a robust detection of the global 21-cm signal, one has to overcome several observational challenges. First, the foregrounds are 4-6 orders of magnitude brighter than the 21-cm signal. Galactic synchrotron emission constitutes the largest contribution to these foregrounds, which is well described by a power law in frequency i.e.\ $T_{\rm sync} \propto \nu^{-\alpha}$ \citep{1999A&A...345..380S}. For separating the 21-cm signal from the foregrounds, the key is to exploit the spectral differences between them. For example, foregrounds can be considered spectrally smooth in contrast with the cosmological 21-cm signal. However, the gain of the observing antenna has a spatial as well as a spectral variation which mixes the spatial distribution of the foreground power into the frequency domain and corrupts the intrinsically smooth foregrounds \citep{Vedantham_2013,Bernardi_2015,Spinelli_2021,Anstey_2021}. These distortions can be sufficiently non-smooth to make them difficult to remove from the data using smooth functions such as polynomials. For this reason, a common goal of all global signal experiments is to strive for as achromatic beams as possible, thereby minimizing any potential leakage of the foregrounds into the 21-cm signal.

To circumvent this issue, it has been proposed by \citet{Vedantham_2013, 2014ApJ...793..102S, Tauscher_2018,Rapetti_2020,Tauscher_2020_util,Bassett_2021} to perform the Singular Value Decomposition (SVD) of the beam-weighted foreground modeling set (created by varying the characteristics of the foreground model) to find the optimal basis with which to fit the foregrounds instead of assuming any specific model. In particular, \citet{Hibbard_2020} have shown that in the case of a perfectly achromatic beam, the basis set obtained from the SVD is very similar to polynomials. However, there is a noticeable difference between the two when the beam is chromatic. In the latter case, the basis obtained from SVD provides a better fit to the beam-weighted foregrounds than polynomials.

A similar approach is used to create a 21-cm modeling set and obtain a 21-cm basis set from the SVD. In this approach, observations are jointly fit by the foreground, and 21-cm basis sets to extract the 21-cm signal. The confidence levels on the extracted signal are determined by the noise level of the data and the covariance between the foreground and 21-cm signal modes. \citet{Tauscher_2018} utilized the observations of the four Stokes parameters to reduce this overlap because the foreground modes appear in all polarization channels, whereas the 21-cm modes only appear in stokes $I$. Another strategy for decreasing this overlap could be obtained by utilizing the time dependence of foregrounds by fitting multiple drift-scan spectra simultaneously as described by \citet{Tauscher_2020_util}. This mitigates the overlap between the modes as the foregrounds change as a function of time, but the global 21-cm signal remains the same across multiple drift-scan spectra. This has also been found in the Bayesian forward modeling approach within REACH \citep{https://doi.org/10.48550/arxiv.2210.04707}.

In this paper, our focus is to explore the robustness of the REACH antenna designs for the 21-cm signal extraction. The designs we currently use are not sensitive to the polarization channels, so we can not utilize this strategy of reducing the overlap between the modes, but we do utilize the time dependence of foregrounds to reduce the overlap. REACH will eventually be observing the sky simultaneously with multiple antennas, including a hexagonal dipole and a conical log spiral antenna \citep{1965ITAP...13..488D}.

This paper is organized as follows: In section~\ref{sec:formalism}, we briefly outline the mathematical formalism of the signal extraction. In section~\ref{sec:sims_modset}, we describe the 21-cm signal and foreground simulations and modeling sets for two different foreground models. Section~\ref{sec:results} describes the results for both the models. In section~\ref{sec:summary}, we summarize our findings.

\section{Formalism}
\label{sec:formalism}
This section briefly outlines the mathematical formalism of the 21-cm signal extraction. For more details, we refer the reader to \citet{Tauscher_2018,Tauscher_2020_util}. We consider a mock observation $\bm{y}$, which is composed of the global 21-cm signal ${\bm y}_{21}$, beam-weighted foregrounds ${\bm y}_{\rm FG}$ and a Gaussian random noise ${\bm n}$ with covariance ${\bm C}$.
\begin{equation}
    {\bm y} = {\bm \Psi}_{21}{\bm y}_{21} + {\bm y}_{\rm FG} + {\bm n}\,,
\end{equation}
where $\*\Psi_{21}$ is a matrix that expands the 21-cm signal, which is only a function of frequency and not time, into the full data vector space. For example, when we simultaneously fit the data across multiple time slices (and antennas), the beam-weighted foregrounds $\*y_{\rm FG}$ will change from one time slice (and antenna) to another; however, $\*y_{21}$ will remain the same. In that case, $\*\Psi_{21}$ expands the 21-cm signal across multiple time bins (and antennas).

There are also errors introduced in the data due to imperfect calibration such as additive biases through the receiver's gain and antenna noise. In this work, we assume a well calibrated instrument and ignore these sources of uncertainty. The impact of these instrumental errors will be considered in a future work. These effects to some extent have been taken into account in \citet{Tauscher_2021r}.

To extract $\*y_{21}$ from the data $\*y$, we model the data using the basis vectors (or modes) for the 21-cm signal $\*F_{21}$ and beam-weighted foregrounds $\*F_{\rm FG}$, derived from the SVD of their modeling sets. These modes are further normalized such that $\*F_{21}^T\*\Psi_{21}^T\*C^{-1}\*\Psi_{21}\*F_{21} = \*I$ and $\*F_{\rm FG}^T\*C^{-1}\*F_{\rm FG} = \*I$, where $\*I$ is the identity matrix. The model of our data then takes the form
\begin{equation}
    \mathcal{\*M} (\*x_{21}, \*x_{\rm FG}) = \*\Psi_{21}\*F_{21}\*x_{21} + \*F_{\rm FG}\*x_{\rm FG}\,,
\end{equation}
where $\*x_{21}$ and $\*x_{\rm FG}$ are the linear coefficients that scale the basis vector amplitudes contained in $\*F_{21}$ and $\*F_{\rm FG}$ respectively. This can be re-written as
\begin{equation}
    \label{eq:model}
    \mathcal{\*M}(\*x) = \*F\*x\,,
\end{equation}
where $\*F$ = $[\*\Psi_{21}\*F_{21}\ |\ \*F_{\rm FG}]$ and $\*x^T = [\*x_{21}^T\ |\ \*x_{\rm FG}^T]$.

To find the parameters $\*x$ of our model $\mathcal{\*M}$, we assume the likelihood to be Gaussian i.e.
\begin{equation}
    \mathcal{L}(\*y|\*x) \propto {\rm exp} \left\{ - \frac{1}{2} [\*y - \*F\*x]^T \*C^{-1} [\*y - \*F\*x] \right\}\,.
\end{equation}
Assuming a flat prior on the scaling parameters\footnote{In principle, one can derive informative priors from the modeling sets; however, it is based on a few key assumptions which limit its applicability (see Appendix C of \citet{Tauscher_2018}).}, the posterior distribution also turns out to be Gaussian in $\*x$ with mean $\*\xi$ and covariance $\*S$ given as
\begin{equation}
    \label{eq:mean_cov}
    \begin{aligned}
    \*\xi &= \*S\*F^T\*C^{-1}\*y\,,\\
    \*S &= (\*F^T\*C^{-1}\*F)^{-1}\,.\\
    \end{aligned}
\end{equation}

The maximum likelihood reconstruction of the 21-cm signal $\*\gamma_{21}$ and its covariance $\*\Delta_{21}$ is given as [see equation~(\ref{eq:model})]
\begin{equation}
    \label{eq:MLE_cov}
    \begin{aligned}
    \*\gamma_{21} &= \*F_{21}\*\xi_{21}\,,\\
    \*\Delta_{21} &= \*F_{21}\*S_{21}\*F_{21}^T\,,
    \end{aligned}
\end{equation}
where $\*\xi_{21}$ and $\*S_{21}$ are part of the $\*\xi$ and $\*S$ matrices respectively, corresponding to the 21-cm component. The 1-$\sigma$ Root-Mean-Square (RMS) uncertainty on the reconstructed signal is then given by 
\begin{equation}
    \label{eq:RMS_1sigma}
    {\rm RMS}_{21}^{1\sigma} = \sqrt{{\rm Tr}(\*\Delta_{21})/n_{\nu}}\,,
\end{equation}
where $n_\nu$ is the number of frequency channels.
\begin{figure}
    \centering
    \includegraphics[width=\linewidth]{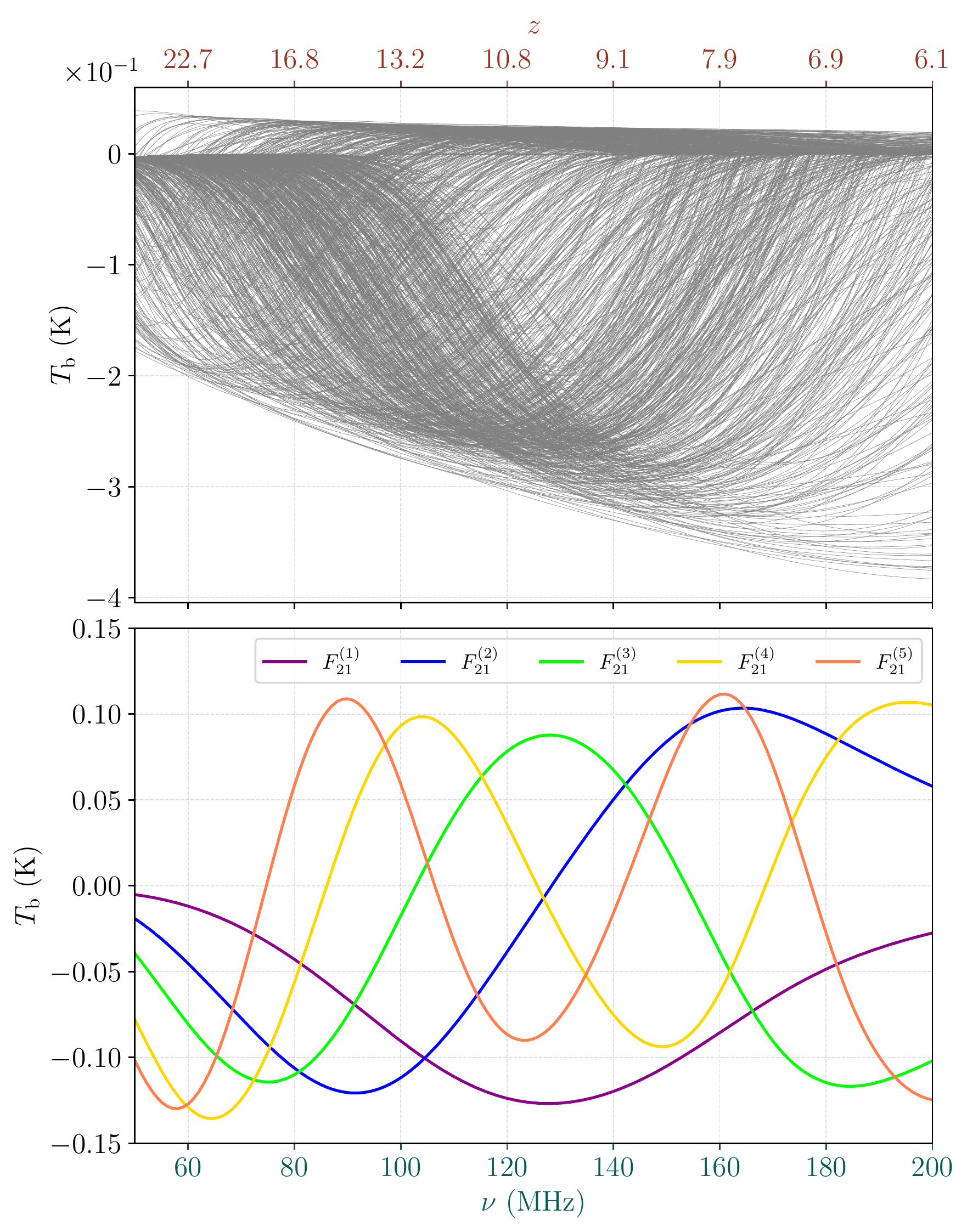}
    \caption{\textbf{Top panel}: A thin slice of the 21-cm signal modeling set composed of all the physical signals simulated using ARES. \textbf{Bottom panel:} The first five modes (purple, blue, green, yellow, orange) obtained from the Singular Value Decomposition of the modeling set.}
    \label{fig:21_modset_basis}
\end{figure}

\begin{figure}
    \centering
    \includegraphics[width=0.8\linewidth]{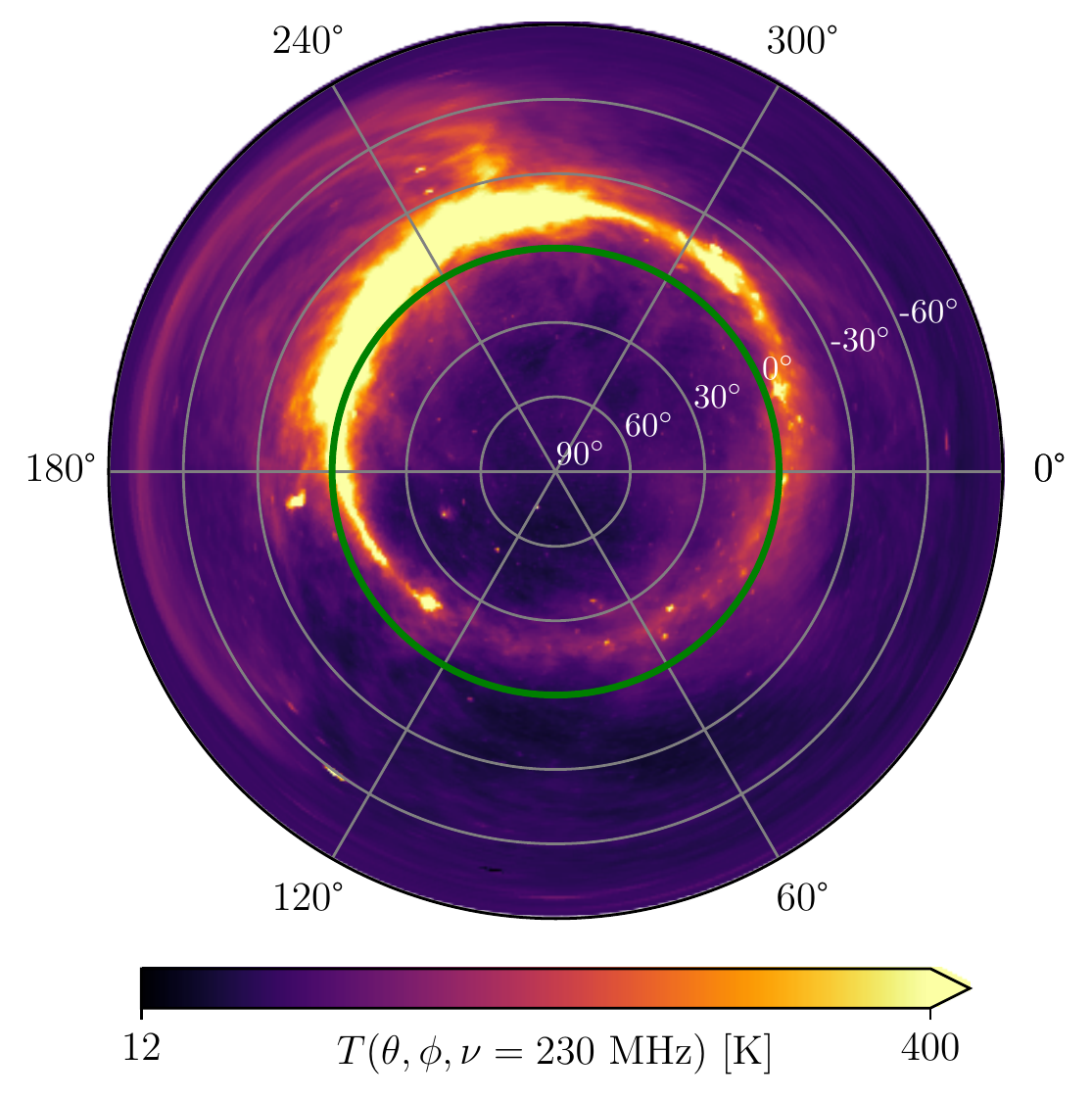}
    \caption{Radio sky \citep{2008MNRAS.388..247D} at 230 MHz for the simulated observation on 2019-10-01 at 00:00:00 UTC when the galactic disk remains below the horizon ($\theta < 0^{\circ}$) for an antenna located in the Karoo radio reserve. This is shown in altitude ($\theta$) and azimuth ($\phi$) coordinates of the antenna's frame with the zenith in the center, where $\theta=0^{\circ}$ (green line) marks the horizon.}
    \label{fig:gsm_map}
\end{figure}

\begin{figure*}
    \centering
    \includegraphics[width=0.875\linewidth]{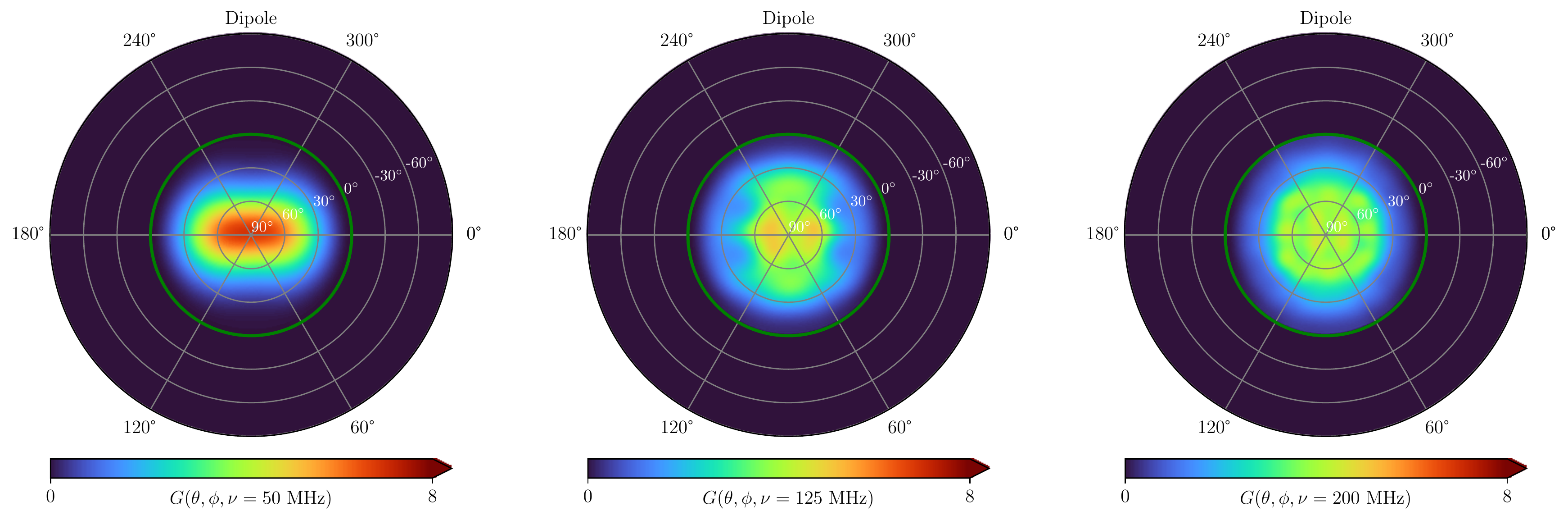}\vspace{0.2em}
    \includegraphics[width=0.875\linewidth]{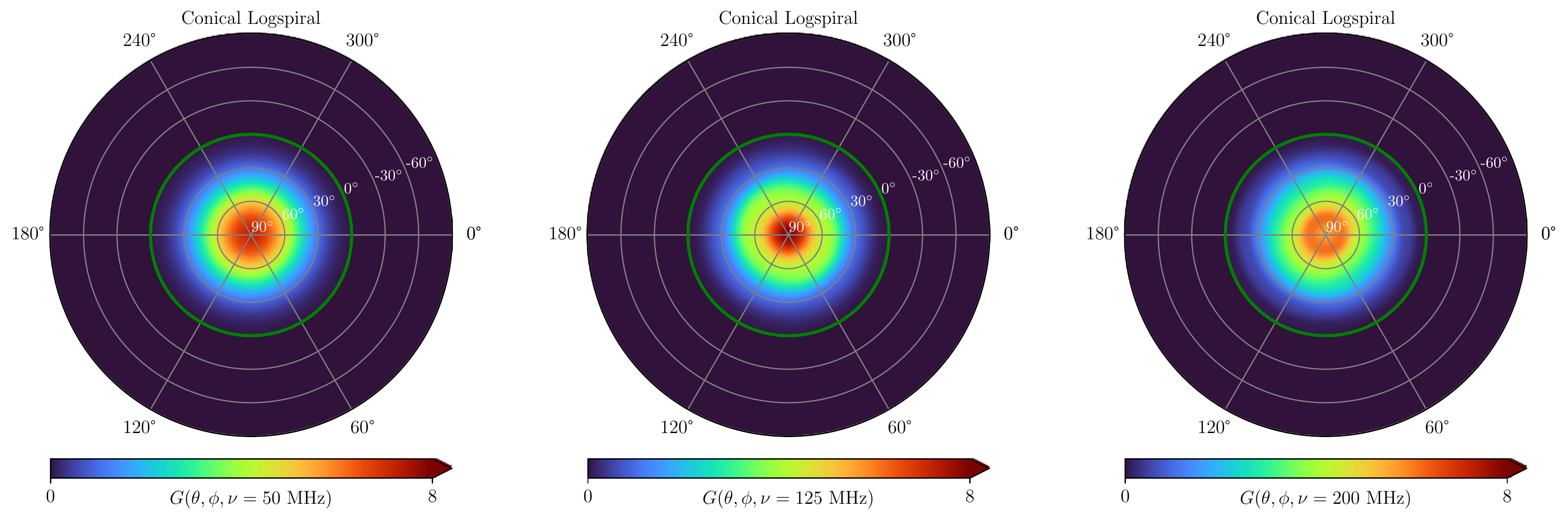}\vspace{0.2em}
    \includegraphics[width=0.875\linewidth]{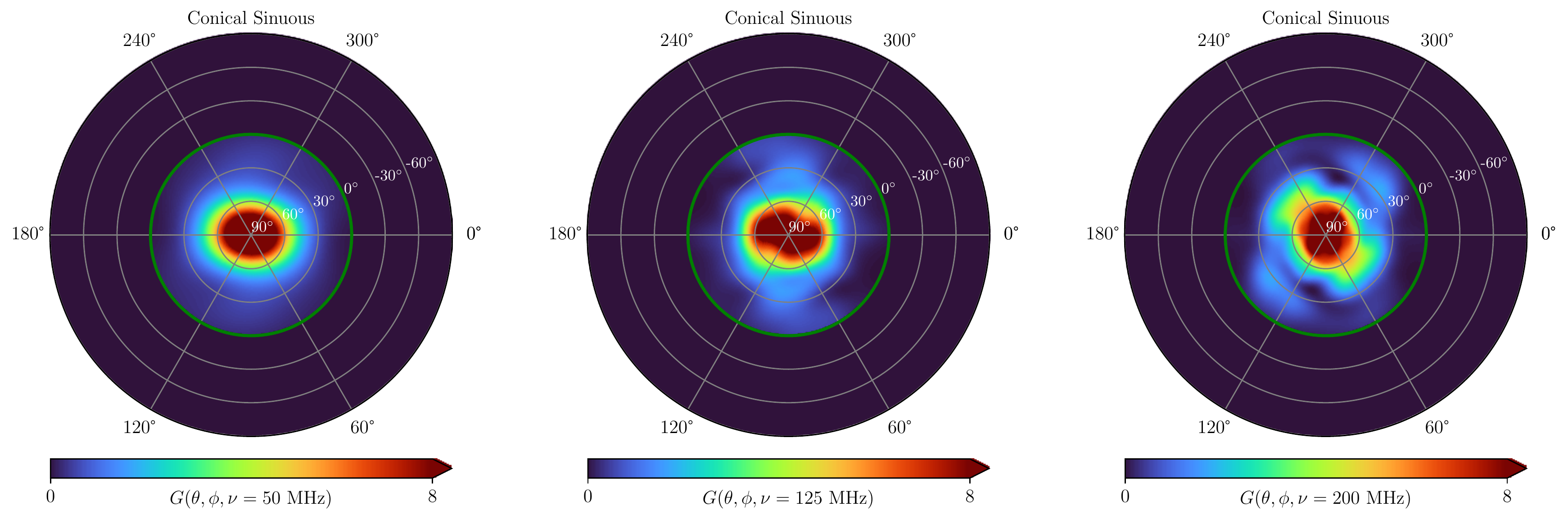}
    \caption{Beam patterns $G(\theta, \phi, \nu)$ for the dipole (first row), conical log spiral (second row), and conical sinuous (third row) antenna at $\nu$ = 50 MHz (first column), 125 MHz (second column) and 200 MHz (third column). These are shown in the altitude ($\theta$) and azimuth ($\phi$) coordinates of the antenna's reference frame with the zenith in the center. The green line ($\theta=0^{\circ}$) marks the horizon.}
    \label{fig:beam_patterns}
\end{figure*}

While fitting the data, we select the number of modes for each component by minimizing the Deviance Information Criterion (DIC) \citep{RePEc:bla:jorssb:v:64:y:2002:i:4:p:583-639}, which is given as
\begin{equation}
    \label{eq:dic}
    {\rm DIC} = \*\delta^T\*C^{-1}\*\delta + 2(n_b^{21} + n_b^{\rm FG})\,,
\end{equation}
where $\*\delta = \*F\*\xi - \*y$, $n_b^{21}$ and $n_b^{\rm FG}$ are the number of 21-cm and foreground modes used in the fit. The first term in equation~(\ref{eq:dic}) represents the goodness of fit and the second term quantifies the model complexity. We also compare its performance with the Bayesian Information Criterion (BIC) and find that minimizing DIC results in more unbiased fits compared to the BIC. This is consistent with \citet{Tauscher_2018}.
\begin{figure}
    \centering
    \includegraphics[width=\linewidth]{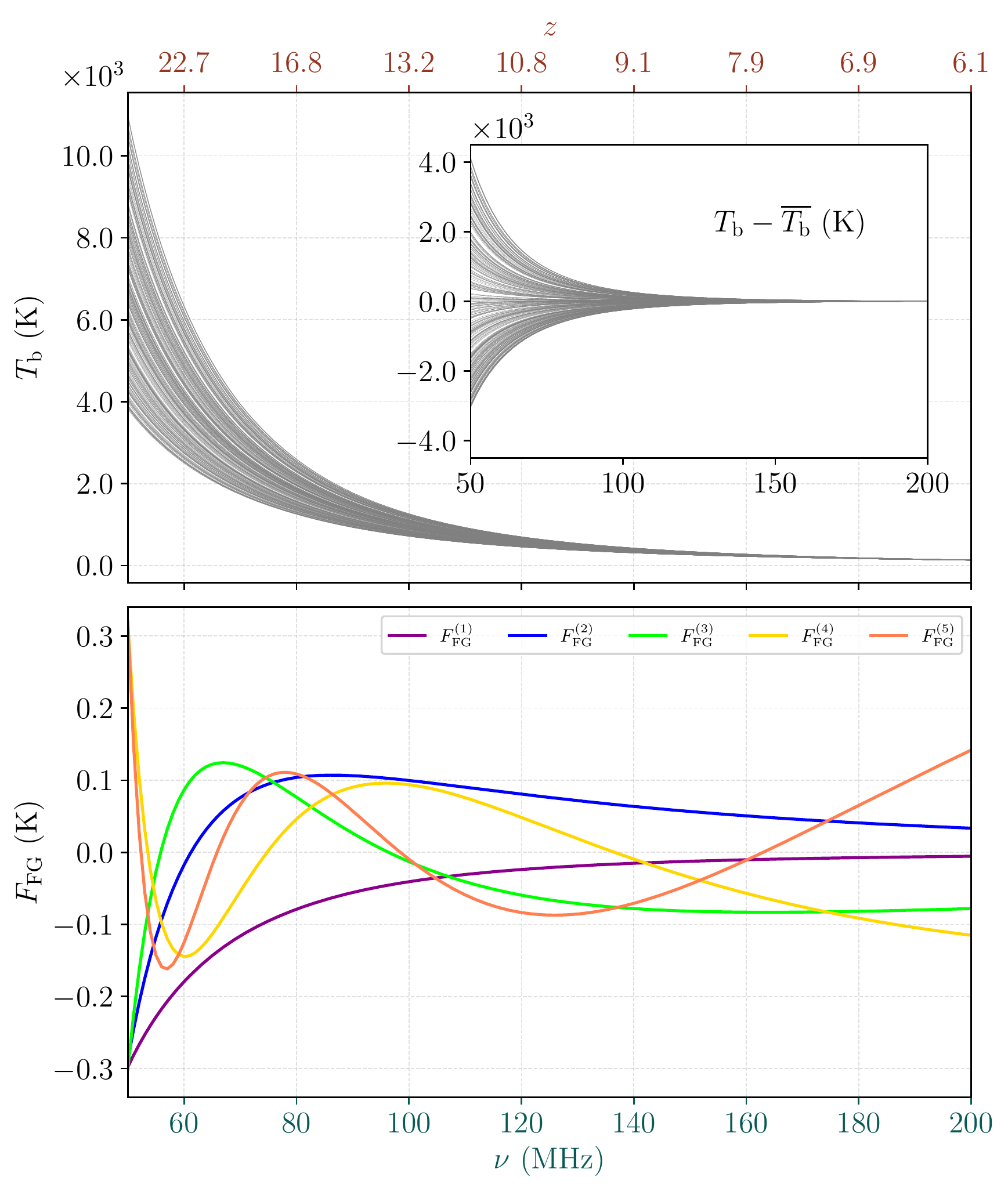}
    \caption{\textbf{Top panel:} A thin slice of the simple beam-weighted foreground modeling set for the dipole antenna. The inset plot shows the same slice of the modeling set with its mean subtracted. \textbf{Bottom panel}: The first five modes (shown by different colors) obtained from the SVD. The sign of the basis function is irrelevant since it is absorbed in the weights.}
    \label{fig:Fg_modset_basis}
\end{figure}

\subsection{Statistical Measures}
\label{subsec:stat_meas}
To examine and compare the performance of different combinations of antennas, we estimate a few statistical measures. First, we estimate the signal bias statistic defined in \citet{Tauscher_2018} as
\begin{equation}
    \label{eq:signal_bias}
    \varepsilon = \sqrt{\frac{1}{n_\nu} \sum_{i=1}^{n_\nu} \frac{(\*\gamma_{21} - \*y_{21})^{2}}{(\*\Delta_{21})_{ii}}}\,,
\end{equation}
where $n_{\nu}$ is the number of frequency channels, $\*y_{21}$ is the input signal, and $\*\gamma_{21}$ and $\*\Delta_{21}$ are the reconstructed signal and its channel covariance respectively from equation~(\ref{eq:MLE_cov}). It determines the number of $\sigma$ at which the uncertainty interval on the extracted signal includes the input 21-cm signal. Now we can define the uncertainty interval, which includes the input signal as
\begin{equation}
    {\rm RMS}_{21} = \varepsilon\,{\rm RMS}_{21}^{1\sigma}\,,
\end{equation}
where $\varepsilon$ and ${\rm RMS}_{21}^{1\sigma}$ are given from equation~(\ref{eq:signal_bias}) and (\ref{eq:RMS_1sigma}) respectively.

The signal bias statistic and RMS uncertainty of the extracted 21-cm signal can only be estimated if the input 21-cm signal is known, which is not the case in practice. So, in order to quantify the overall quality of the fit to the full data $\*y$, we calculate the normalized chi-squared statistic as
\begin{equation}
    \label{eq:norm_chi2}
    \chi^2 = \frac{\*\delta^T \*C^{-1} \*\delta}{N_{c} - N_{p}}\,,
\end{equation}
where $N_c$ is the number of data channels, and $N_p$ is the number of parameters (or modes) used in the fit.

\section{Simulations and modeling sets}
\label{sec:sims_modset}
We begin by creating the modeling sets for the 21-cm signal and the beam weighted foregrounds. For the 21-cm signal, we form a modeling set from all the physical signals simulated using the code Accelerated Reionization Era Simulations (\texttt{ares})\footnote{\url{https://ares.readthedocs.io/en/v0.8/}} \citep{Mirocha_2015, 2017MNRAS.464.1365M}. In the top panel of Figure~\ref{fig:21_modset_basis}, we show a subset of the entire 21-cm modeling set. We then perform the SVD of this modeling set to derive the modes of variation which form an orthonormal basis. The first five modes are shown in the bottom panel of Figure~\ref{fig:21_modset_basis}. These modes are numbered (from violet to orange) according to their ability to capture the variation in the modeling set.

A foreground model is derived from the following features: (i) a spatial brightness temperature distribution, (ii) a spatial spectral index distribution, and (iii) the frequency-dependent beam pattern of the observing antenna. For simulating such beam-weighted foregrounds, we consider two different models based on their complexity.

\subsection{A Simple Foreground Model}
\label{sec:simple_fgmodel}
In our simplest foreground model, we assume the spectral index to be constant across the sky. We generate our sky model as
\begin{equation}
    \label{eq:sky}
    T_{\rm sky} (\theta, \phi, \nu, t) = \left( T_{230}(\theta, \phi, t) - T_{\rm CMB} \right) \left( \frac{\nu}{230} \right)^{-\beta} + T_{\rm CMB}\,,
\end{equation}
where $T_{230} (\theta, \phi, t)$ is the spatial brightness temperature distribution derived from the Global Sky Map at 230 MHz \citep{2008MNRAS.388..247D}, which we use as our base foreground map, and $T_{\rm CMB} = 2.725$ K. The choice of our foreground map follows from an independent Bayesian analysis of REACH \citep{https://doi.org/10.48550/arxiv.2210.04707}; however, a physical sky model such as GMOSS \citep{Rao_2016} can also be used which encapsulates the effects of radiative processes from different components of the radio sky. Our simulated observations start at 2019-10-01 00:00:00 UTC and are integrated for 6 hours. In Figure~\ref{fig:gsm_map}, we show the overhead sky at 230 MHz on 2019-10-01 00:00:00 UTC for an antenna located in the Karoo radio reserve in South Africa.

We convolve this sky model with the beam pattern of the antenna to form the beam-weighted foregrounds as
\begin{equation}
    \label{eq:bwfg}
    T_{\rm FG}(\nu, t) = \frac{1}{4\pi} \int_0^{4\pi} G(\theta, \phi, \nu)\ T_{\rm sky} (\theta, \phi, \nu, t)\, {\rm d}\Omega\,,
\end{equation}
where $G(\theta, \phi, \nu)$ is the beam pattern of the antenna with the beam centered on the zenith. Within REACH, we consider here three different antenna designs: a dipole, a conical sinuous, and a conical log spiral antenna. In Figure~\ref{fig:beam_patterns}, we show the beam patterns $G(\theta, \phi)$ of these antennas at three different frequencies $\nu$ = 50, 125, and 200 MHz in the antenna's reference frame. The distortions caused by the chromaticity of these antennas are sufficiently non-smooth to prevent proper fitting by smooth polynomial-based foreground models, thereby masking the 21-cm signal (see Figure 5 of \citet{Anstey_2021}). Our analysis is based on the observations during which the galactic plane remains below the horizon, as these distortions are more pronounced when the galaxy is above the horizon.

\begin{figure*}
    \centering
    \includegraphics[width=0.34\linewidth]{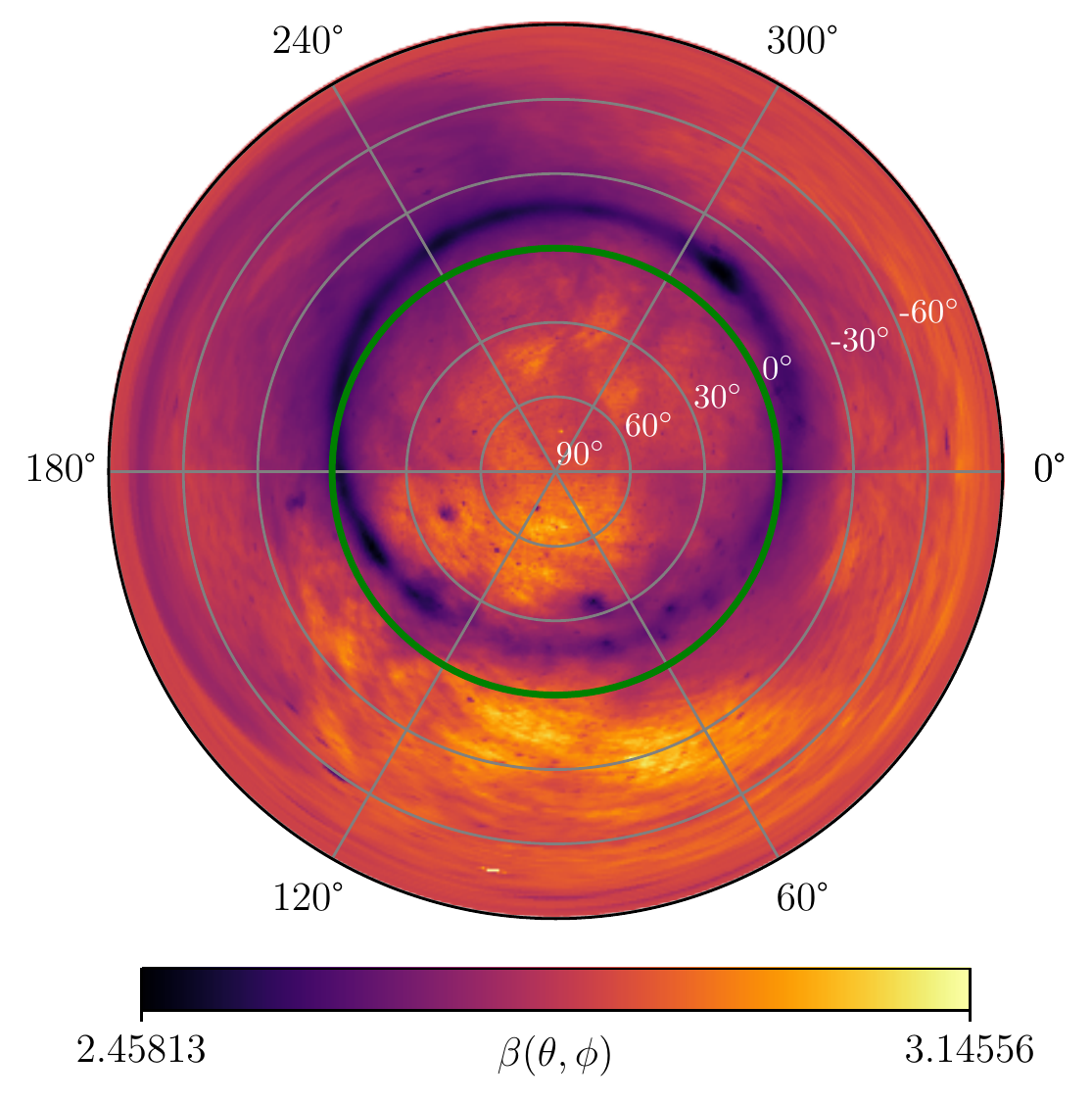}\hspace{4em}
    \includegraphics[width=0.34\linewidth]{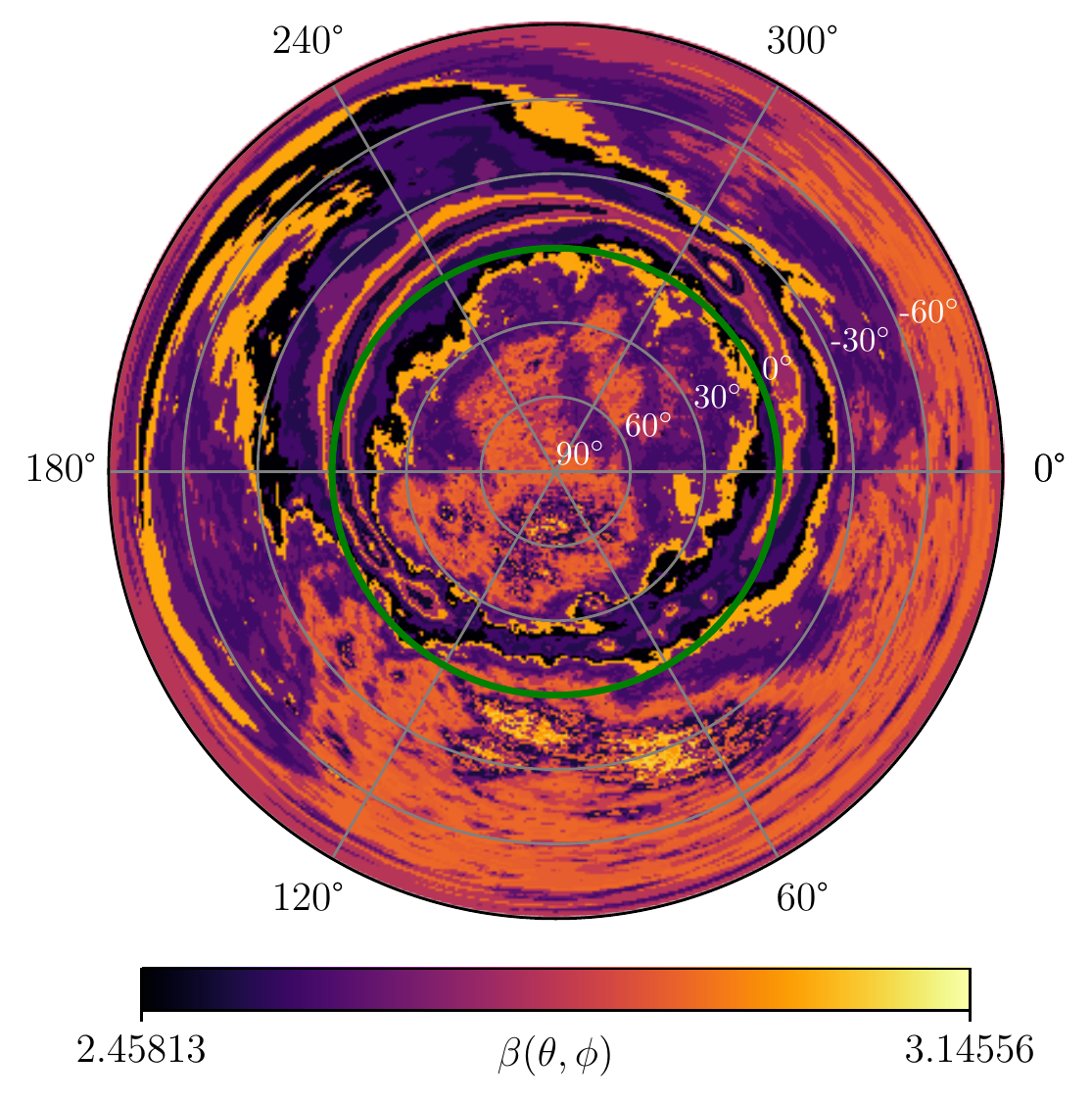}
    \caption{\textbf{Left panel:} Spectral index distribution on 2019-10-01 at 00:00:00 UTC in the altitude ($\theta$) and azimuth ($\phi$) coordinates of the antenna's reference frame, derived from the pixel-wise tracing between two instances of 2008 Global Sky Map at 408 MHz and 230 MHz, \textbf{Right panel:} The division of the sky into 30 regions where the spectral index in each region is randomly sampled from $\beta_i \in (2.45813, 3.14556)$ for $1 \leq i \leq 30$. The green line ($\theta=0^{\circ}$) marks the horizon.}
    \label{fig:spectral_map_division}
\end{figure*}

\begin{figure}
    \centering
    \includegraphics[width=\linewidth]{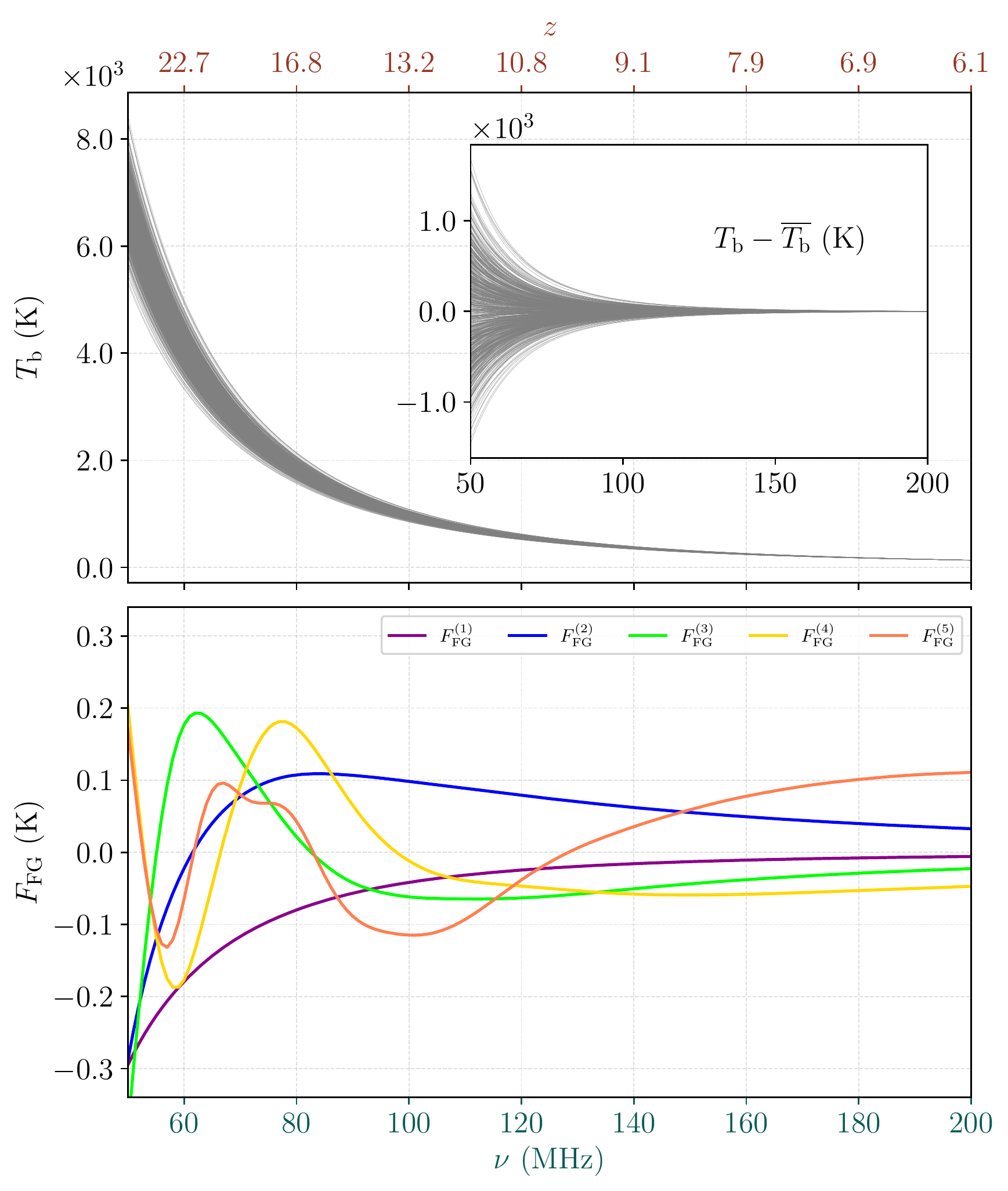}    
    \caption{\textbf{Top panel:} A thin slice of the detailed beam-weighted foreground modeling set for the dipole antenna. The inset plot shows the same slice of the modeling set with its mean subtracted. \textbf{Bottom panel}: The first five modes (shown by different colors) obtained from the SVD.}
    \label{fig:dFg_modset_basis}
\end{figure}

\begin{figure}
    \centering
    \includegraphics[width=0.93\linewidth]{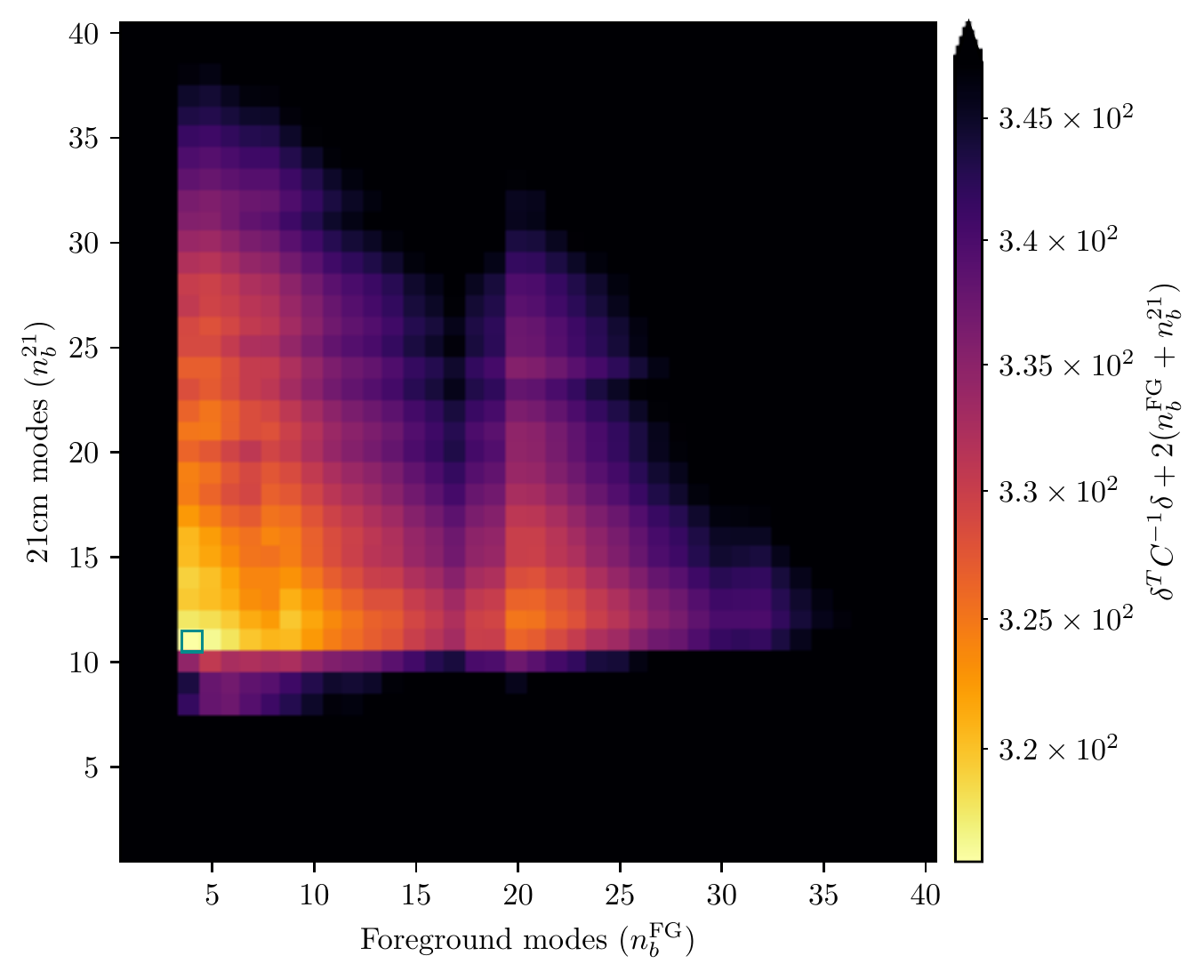}
    \caption{DIC minimization: The colorbar represents the DIC value estimated over a grid of the number of foreground and 21-cm modes. In this case, DIC is minimized with $(n_{b}^{\rm FG}, n_b^{21}) = (4, 11)$.}
    \label{fig:DIC_minimization}
\end{figure}

To form the beam-weighted foreground modeling set for each antenna, we use the spectral index $\beta$ as the free parameter, which is varied in the range $(2.45, 3.15)$. In the top panel of Figure~\ref{fig:Fg_modset_basis}, we show a subset of the beam-weighted foreground modeling set for the dipole antenna. The first five modes obtained from the SVD are shown in the bottom panel.
\subsection{A Detailed Foreground Model}
\label{subsec:detailed_modset}
So far, for simplicity, we assumed that the spectral index is constant. In general, a spatial variation of the spectral index across the sky is expected. This spectral index variation map is derived by calculating the spectral index required to map each pixel of an instance of 2008 Global Sky Map (GSM) at 408 MHz to the corresponding pixel of an instance at 230 MHz as
\begin{equation}
    \beta(\theta, \phi) = \frac{\log{\left(\frac{T_{230} (\theta, \phi) - T_{\rm CMB}}{T_{408} (\theta, \phi) - T_{\rm CMB}}\right)}}{\log{\left(\frac{230}{408}\right)}}\,.
\end{equation}

In the left panel of Figure~\ref{fig:spectral_map_division}, we show the resulting spectral index map. We use this spectral index map to render our mock observations alongside the GSM as the base foreground map with the beam patterns shown in Figure~\ref{fig:beam_patterns}.

To model the foregrounds in a parameterized way, we divide the sky into 30 regions and assign a constant spectral index to each region \citep{Anstey_2021}. To form these regions, we divide the full range of spectral indices shown in the left panel of Figure~\ref{fig:spectral_map_division} into 30 equal intervals, and each region is defined as the patch of the sky with spectral indices within these intervals. We show the division of the sky in the right panel of Figure~\ref{fig:spectral_map_division}.
\begin{figure*}
    \centering
    \includegraphics[width=0.32\linewidth]{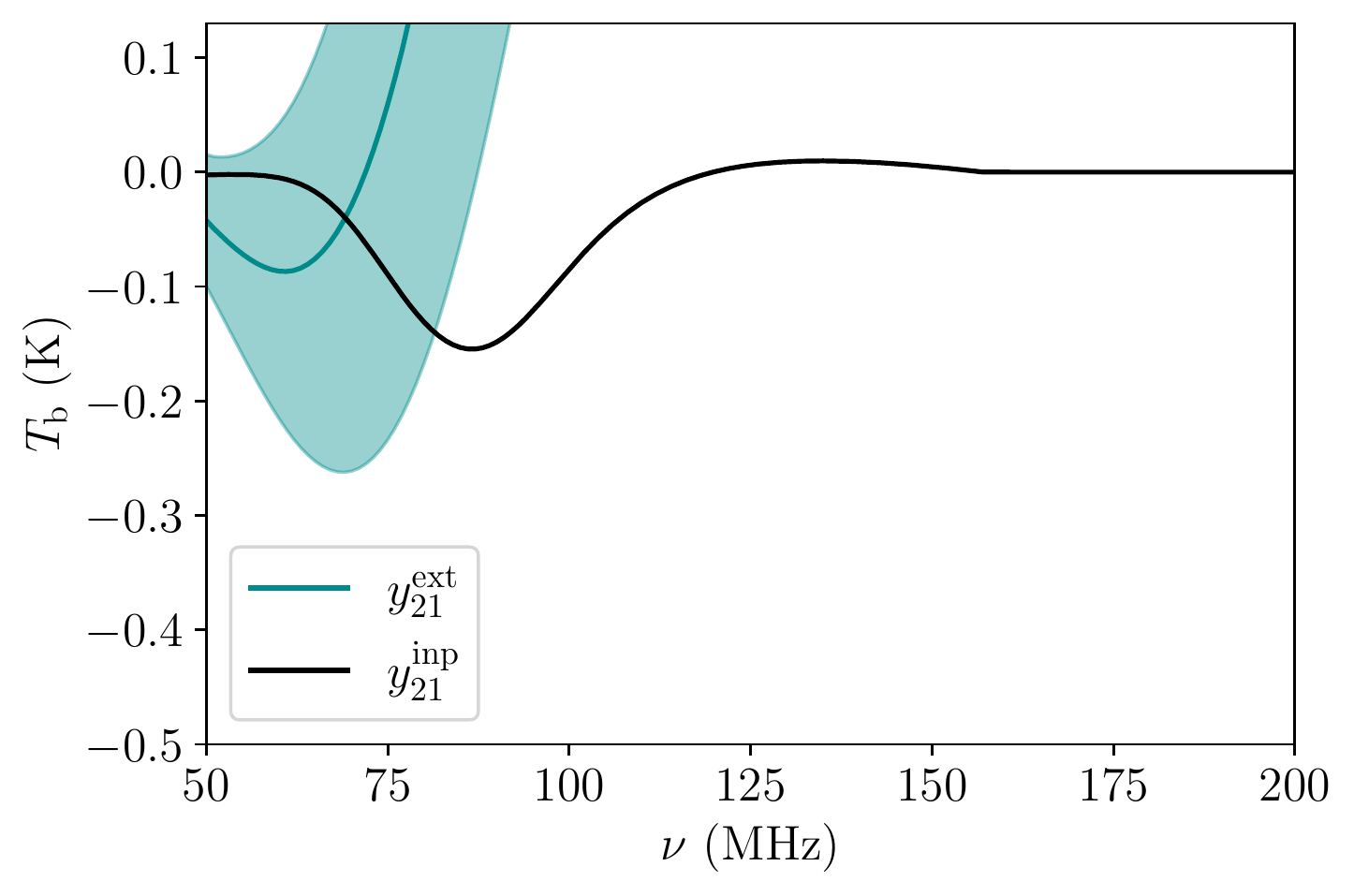}
    \includegraphics[width=0.32\linewidth]{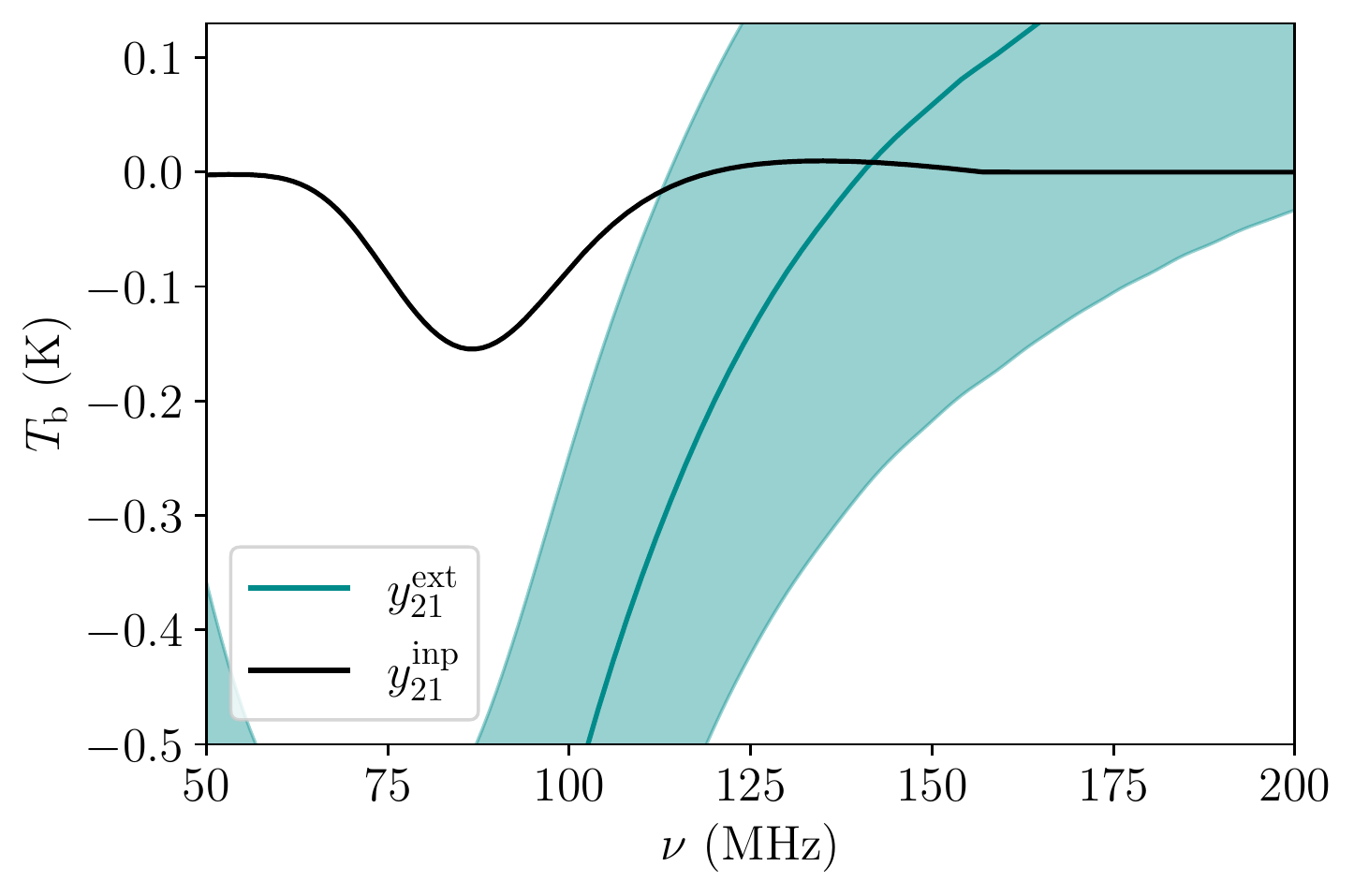}
    \includegraphics[width=0.32\linewidth]{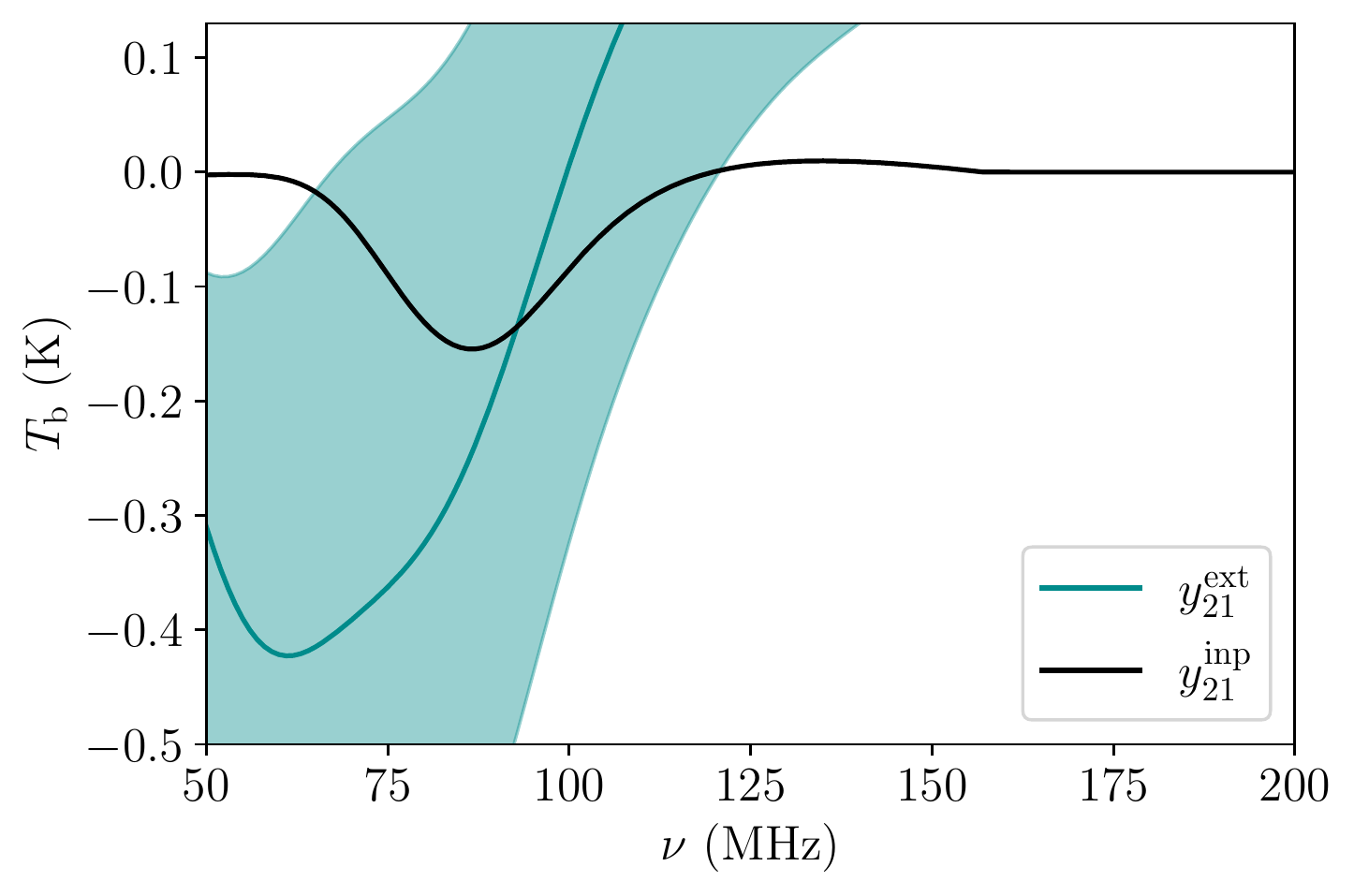}\vspace{0.4em}
    \includegraphics[width=0.32\linewidth]{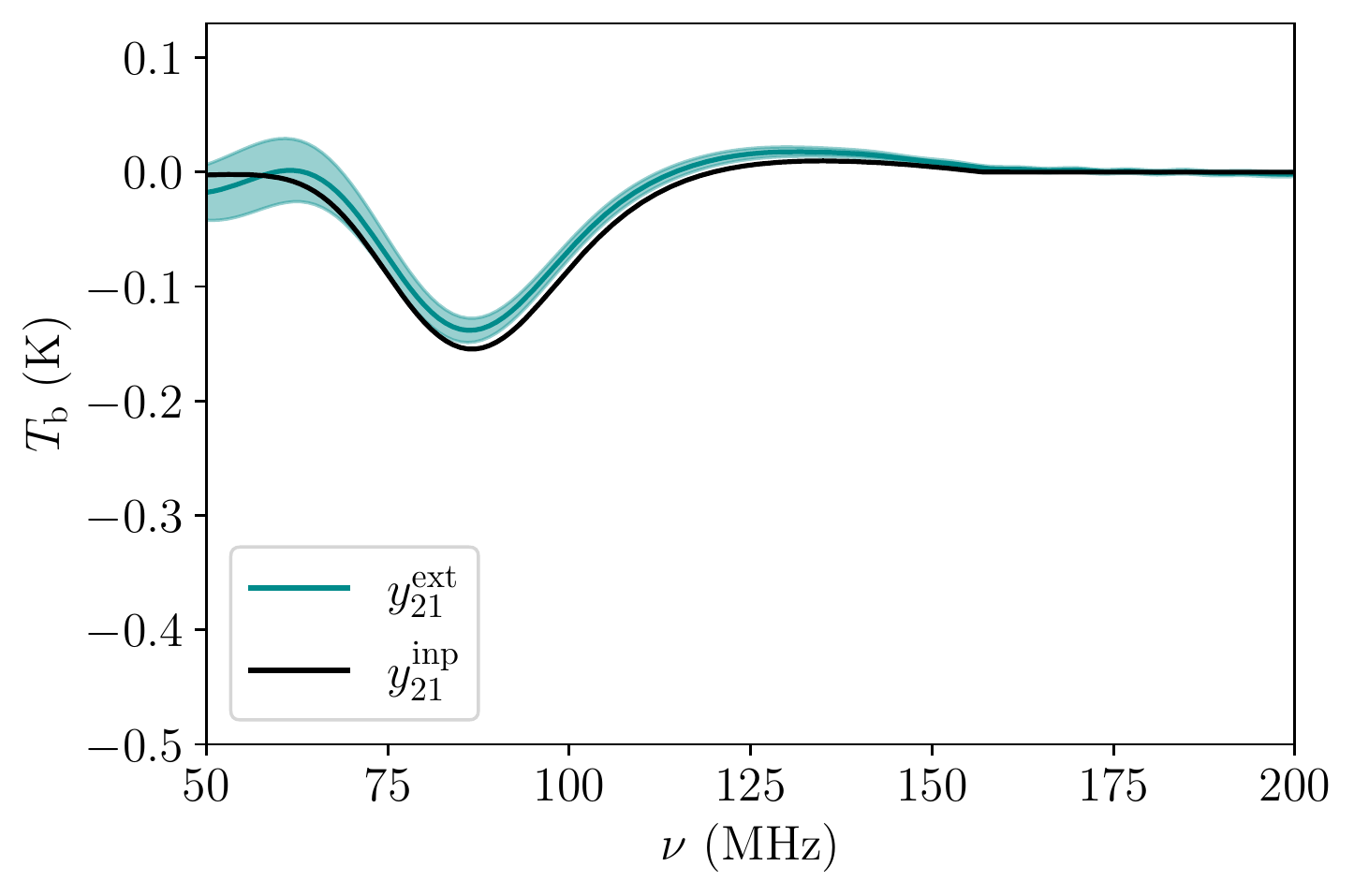}
    \includegraphics[width=0.32\linewidth]{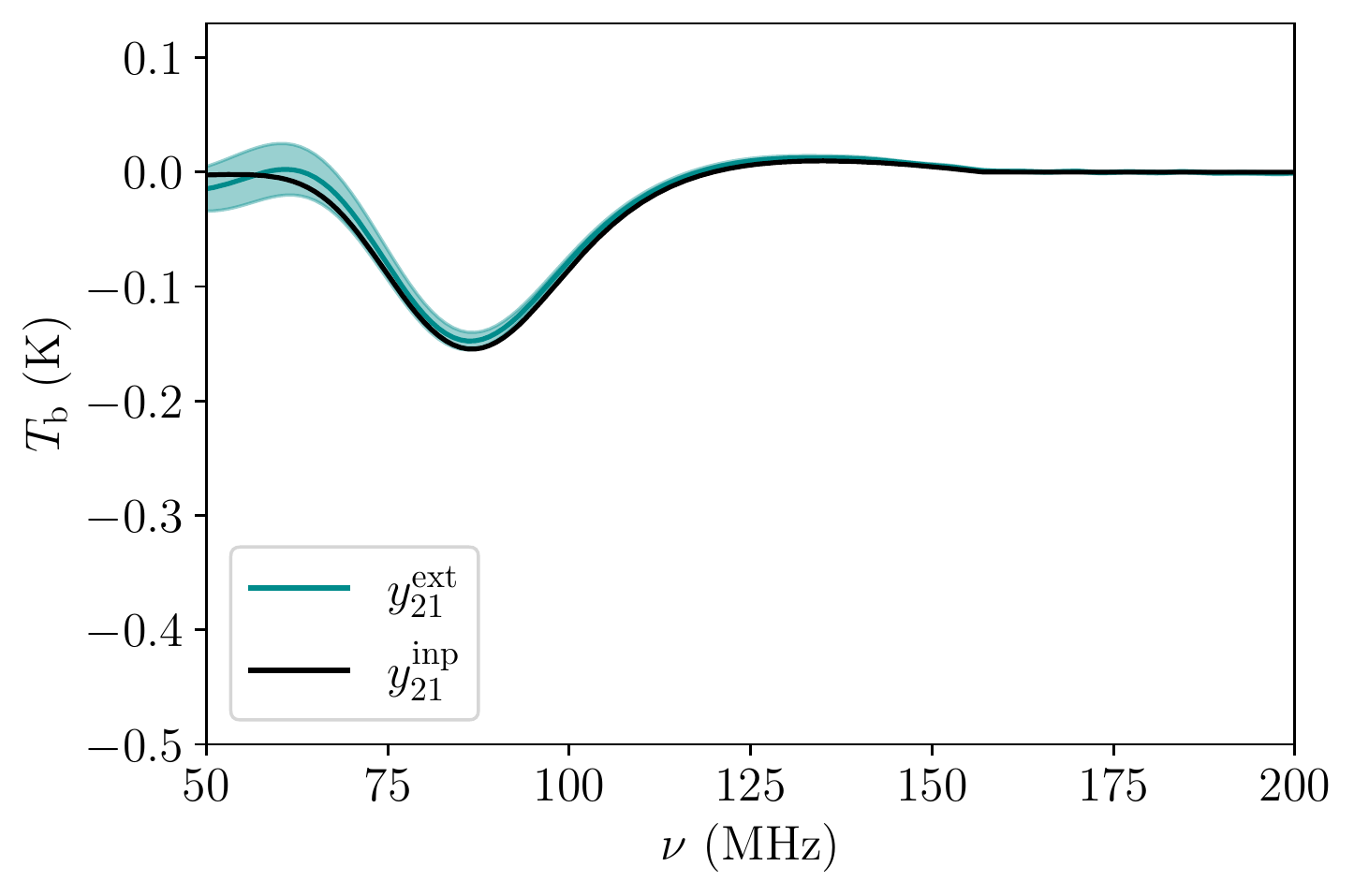}
    \includegraphics[width=0.32\linewidth]{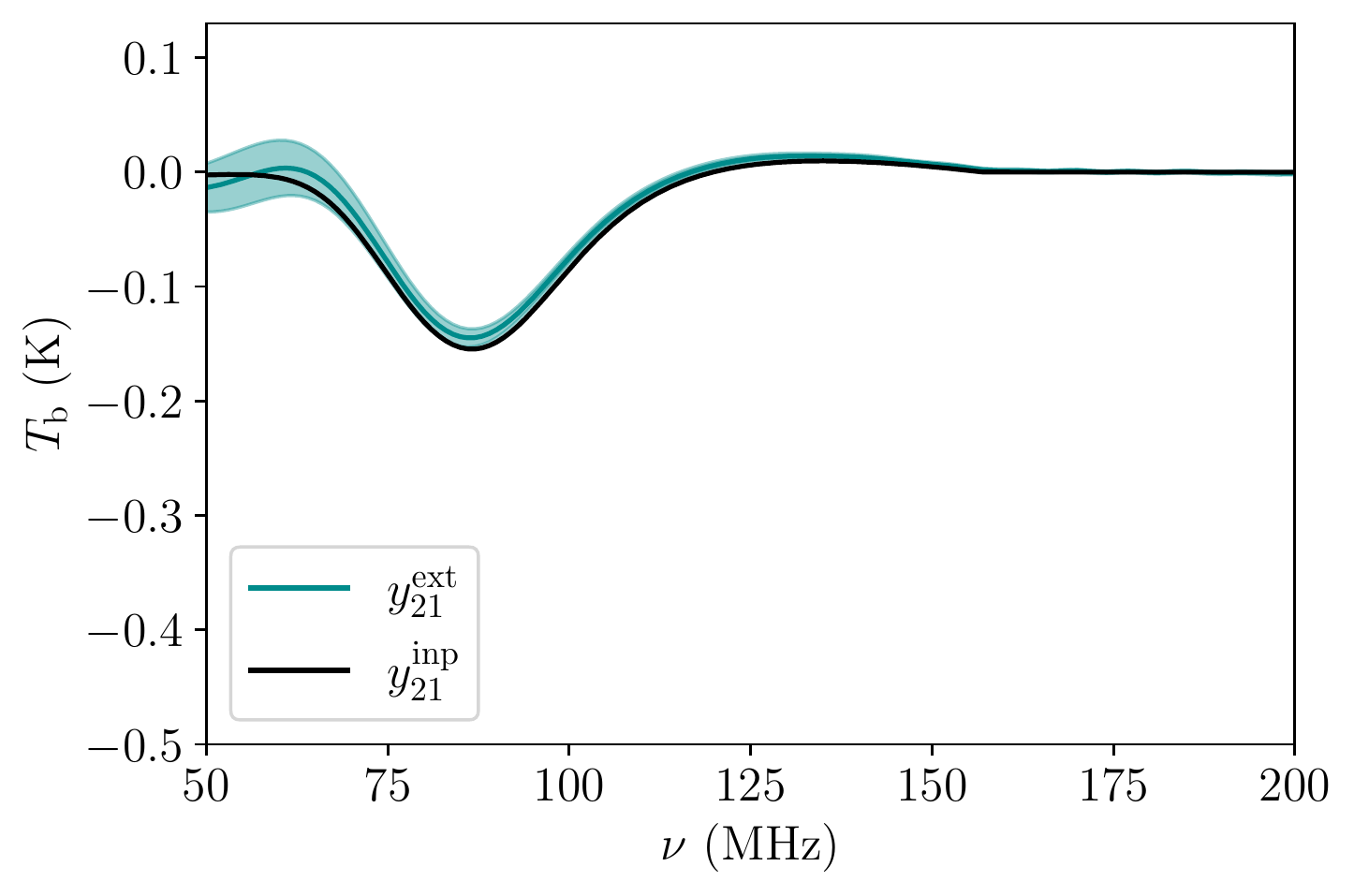}
    \caption{Extracted 21-cm signals for the mock observation generated from a physical 21-cm signal and a foreground model taken from the simple foreground modeling set for dipole (left), conical sinuous (middle), and conical log spiral (right) antenna while simultaneously fitting one time slice (top) and four time slices (bottom). The shaded region represents the 1$\sigma$ confidence region, and the black line shows the input 21-cm signal.}
    \label{fig:ext_nreg_1}
\end{figure*}

To form the beam-weighted foregrounds, we use equation~(\ref{eq:sky}) and (\ref{eq:bwfg}) with $\beta (\theta, \phi)$ as the spectral index model. For creating a modeling set, we randomly sample the spectral index for each region $\beta_i \in (2.45813, 3.14556)$ for $1\leq i\leq30$, which results in different spectral index maps. One of these sample maps is shown in the right panel of Figure~\ref{fig:spectral_map_division}. We take 50,000 samples to form the beam-weighted foreground modeling set. In the top panel of Figure~\ref{fig:dFg_modset_basis}, we show a subset of the modeling set for the dipole antenna. The first five modes obtained from the SVD are shown in the bottom panel.

Once we obtain the modes for the 21-cm signal and the foregrounds, we fit the mock observation by selecting the number of modes that minimize the Deviance Information Criterion. In Figure~\ref{fig:DIC_minimization}, we show this minimization for a typical observation where we estimate the DIC over a grid of 21-cm and foreground modes. In this case, we select four foreground modes and eleven signal modes, as shown with the blue squared box in the figure.

\begin{figure*}
    \centering
    \includegraphics[width=0.46\linewidth]{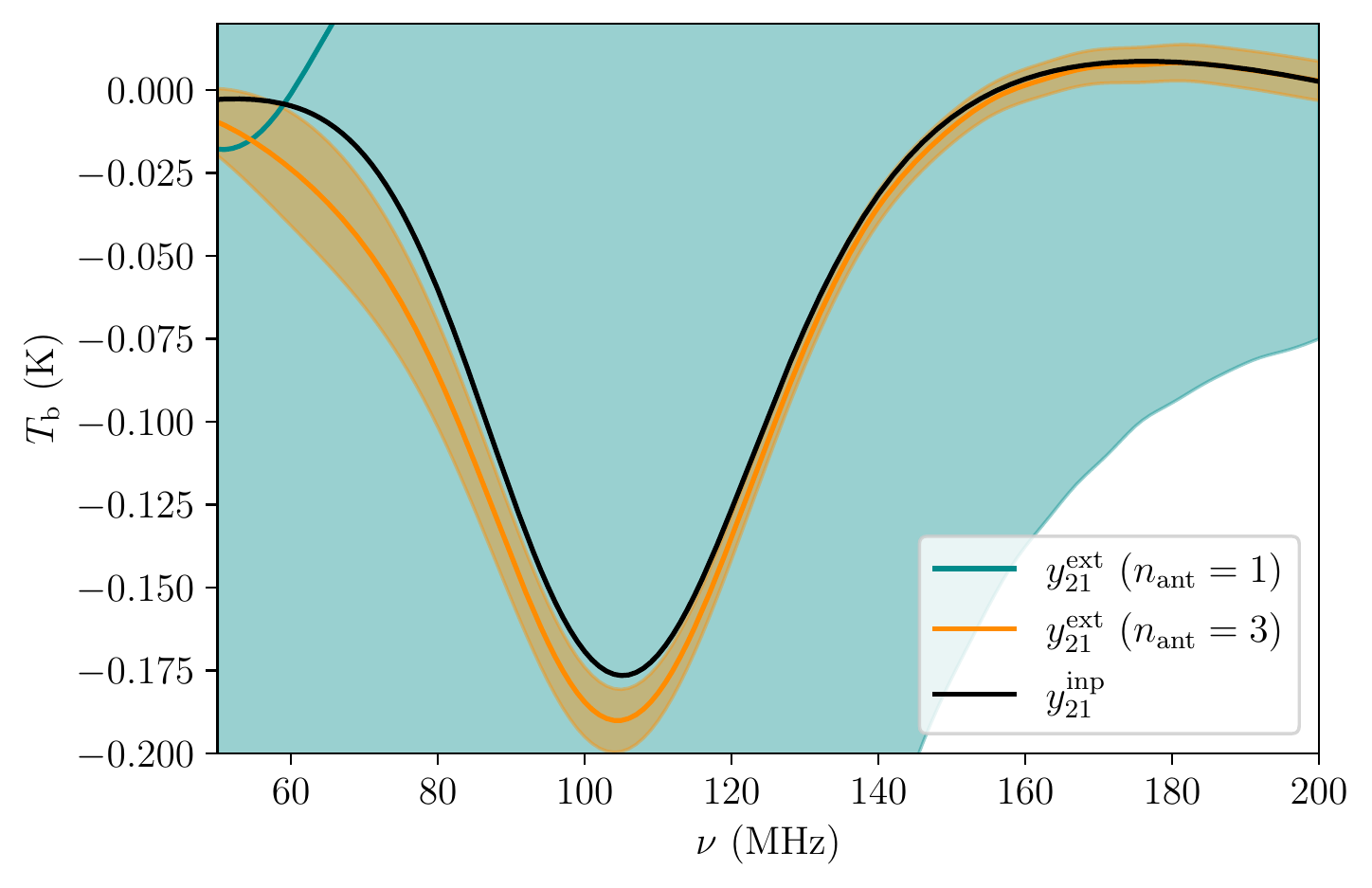}\hspace{2em}
    \includegraphics[width=0.46\linewidth]{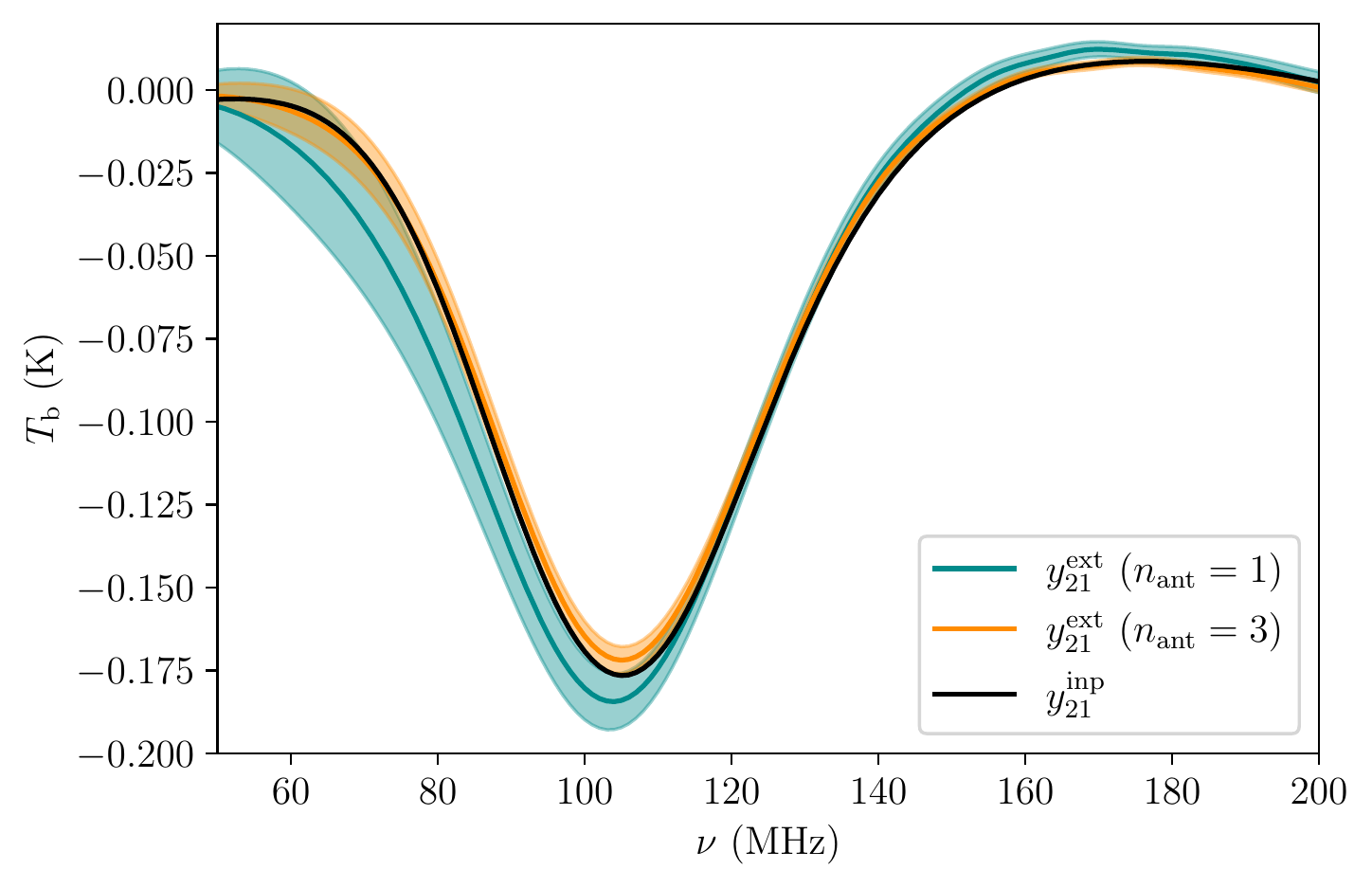}
    \caption{Extracted 21-cm signal for a single antenna (cyan) and multiple antennas (orange) while fitting a single time bin (left panel) and simultaneously fitting two time bins (right panel). The shaded regions represent the 1$\sigma$ confidence region for each case, and the black line shows the input 21-cm signal.}
    \label{fig:ext_ant_comp}
\end{figure*}

\section{Results}
\label{sec:results}
In this section, we discuss the application of this formalism in the 21-cm signal extraction with different antenna designs for the two foreground models that we consider.

\subsection{Simple Foreground Model}
In Figure~\ref{fig:ext_nreg_1}, we show the 21-cm signal extraction for different antennas and the different number of time slices. In each case, the simulated mock observation is generated by taking a 21-cm signal from its modeling set and a foreground component from the simple foreground modeling set (see Section~\ref{sec:simple_fgmodel}). In the first row, we show the extracted signal when we fit the data for a single time slice. The shaded regions represent the 1$\sigma$ confidence intervals. In this case, the pipeline fails to extract the 21-cm signal as the RMS uncertainty for any antenna design remains around 1000 mK. This happens due to a significant covariance between the foreground and 21-cm signal modes. However, we note that in each case, the $\chi^2 \approx 1$, which indicates that the full data gets a good fit by choice of basis vectors from the minimization of the DIC.

To reduce the overlap between the foreground and 21-cm signal modes, we simultaneously fit the data from multiple time slices. While doing so, we utilize the correlations between different time slices and enforce that the 21-cm signal does not change between these slices. In the second row of Figure~\ref{fig:ext_nreg_1}, we show the signal estimates while simultaneously fitting four time bins. We see that the signal extraction significantly improves with an RMS uncertainty for any antenna design around 10 mK. These findings are consistent with the earlier studies by \citet{Tauscher_2020_util}.

\begin{figure}
    \centering
    \includegraphics[width=\linewidth]{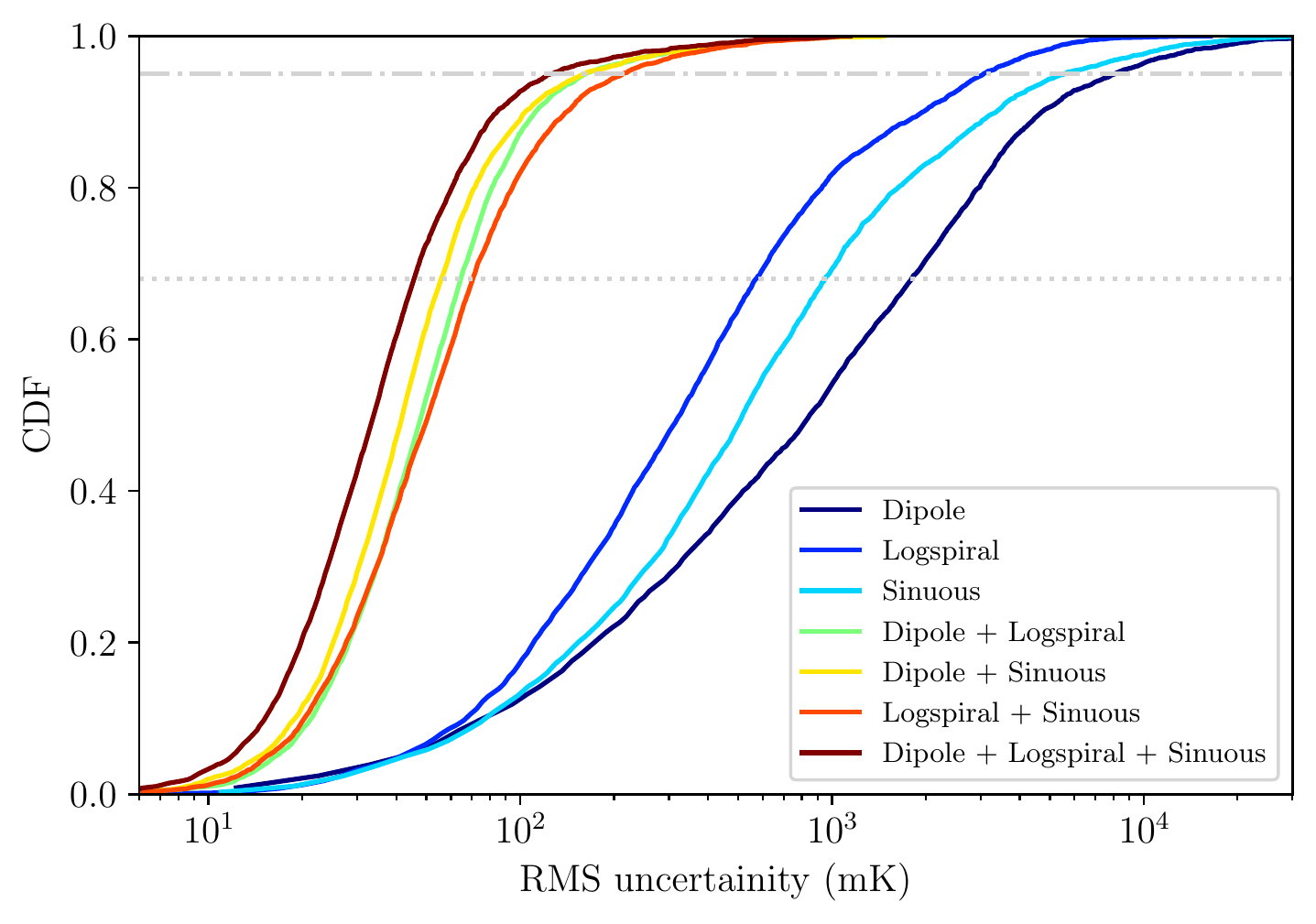}\vspace{0.4em}
    \includegraphics[width=\linewidth]{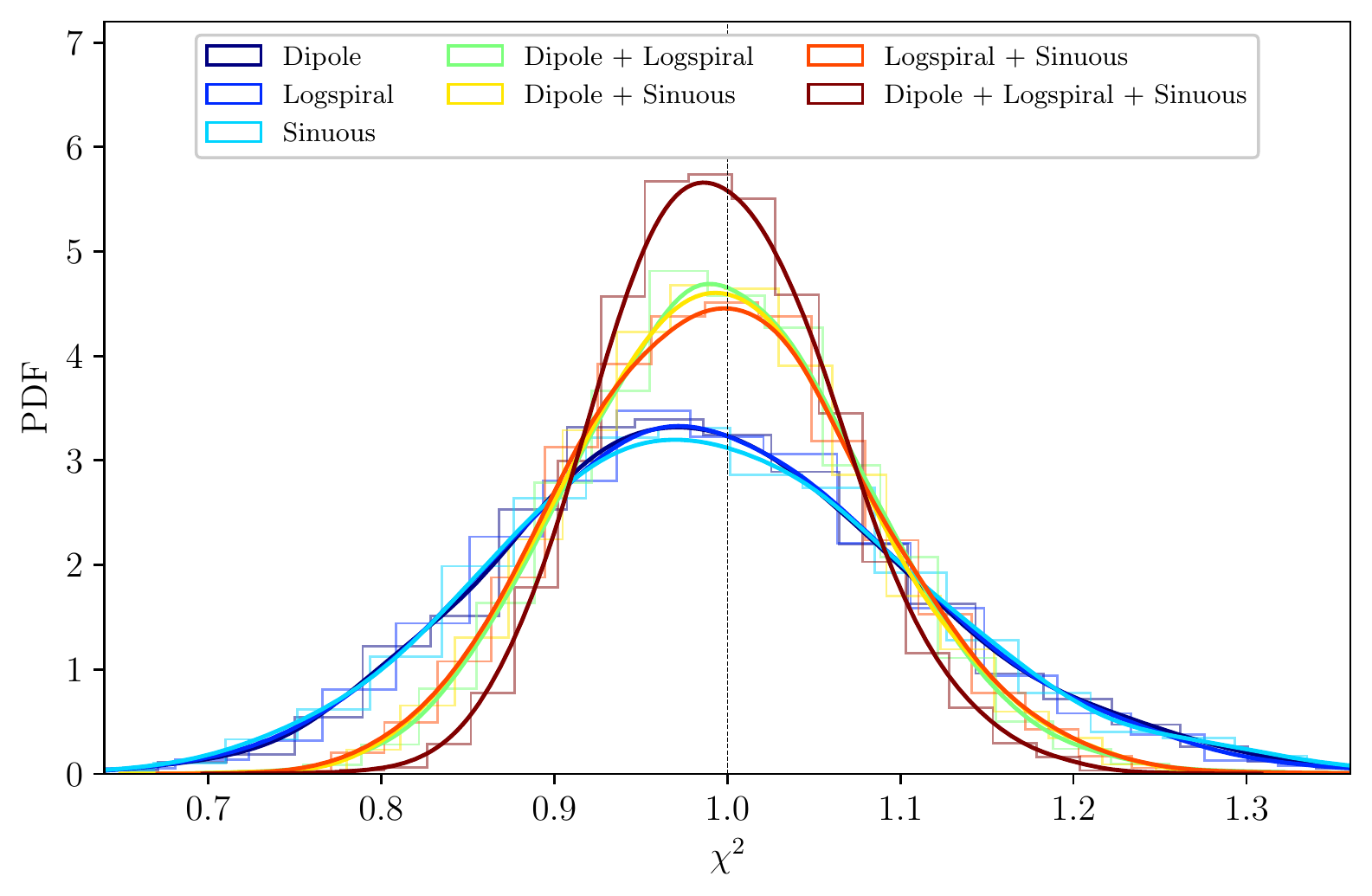}
    \caption{\textbf{Top panel:} Cumulative Distribution Function (CDF) of RMS uncertainty level on the extracted signal estimated from 5000 different mock observations for the simple foreground model. Different colors in the plot represent different combinations of antennas used in the analysis. The uncertainty levels at $68\%$ and $95\%$ for each combination are shown in Table~\ref{tab:tab_simple}. \textbf{Bottom panel:} Probability Distribution Function (PDF) of the normalized $\chi^2$ statistic estimated from 5000 different mock observations for different combinations of antennas.}
    \label{fig:stats_comp}
\end{figure}

Next, we demonstrate how we can utilize the multiple antenna designs within REACH to improve the extraction of the 21-cm signal further. For this, we compare the signal extraction from a single antenna with the extraction from multiple antennas in Figure~\ref{fig:ext_ant_comp}. Note that we assumed a total integration time of 24 hours for both scenarios. In the left panel of Figure~\ref{fig:ext_ant_comp}, we fit the data for a single time slice, where we can clearly see the advantage of simultaneously fitting the data from multiple antennas. In this case, the RMS uncertainty of the single antenna fit (cyan) is 747 mK, which reduces to 23 mK for multiple antennas (orange). This happens because simultaneously fitting multiple antennas utilizes the correlations between different antennas and enforces that the global 21-cm signal does not depend on the antenna design, thereby reducing the covariance between the foreground and 21-cm modes. In the right panel of Figure~\ref{fig:ext_ant_comp}, we simultaneously fit the data from two time slices. In this case, the RMS uncertainty of the single antenna fit (cyan) is 32 mK which reduces to 11 mK for three antennas (orange). Also, in this case, we find that fitting the data from multiple antennas results in a lower uncertainty on the extracted 21-cm signal with a slightly reduced bias, although this improvement is not as pronounced as it is for a single time slice (the left panel). This is because fitting multiple time slices (the right panel) even for a single antenna already removes a significant overlap between the foreground and 21-cm modes.
\begin{figure*}
    \centering
    \includegraphics[width=0.32\linewidth]{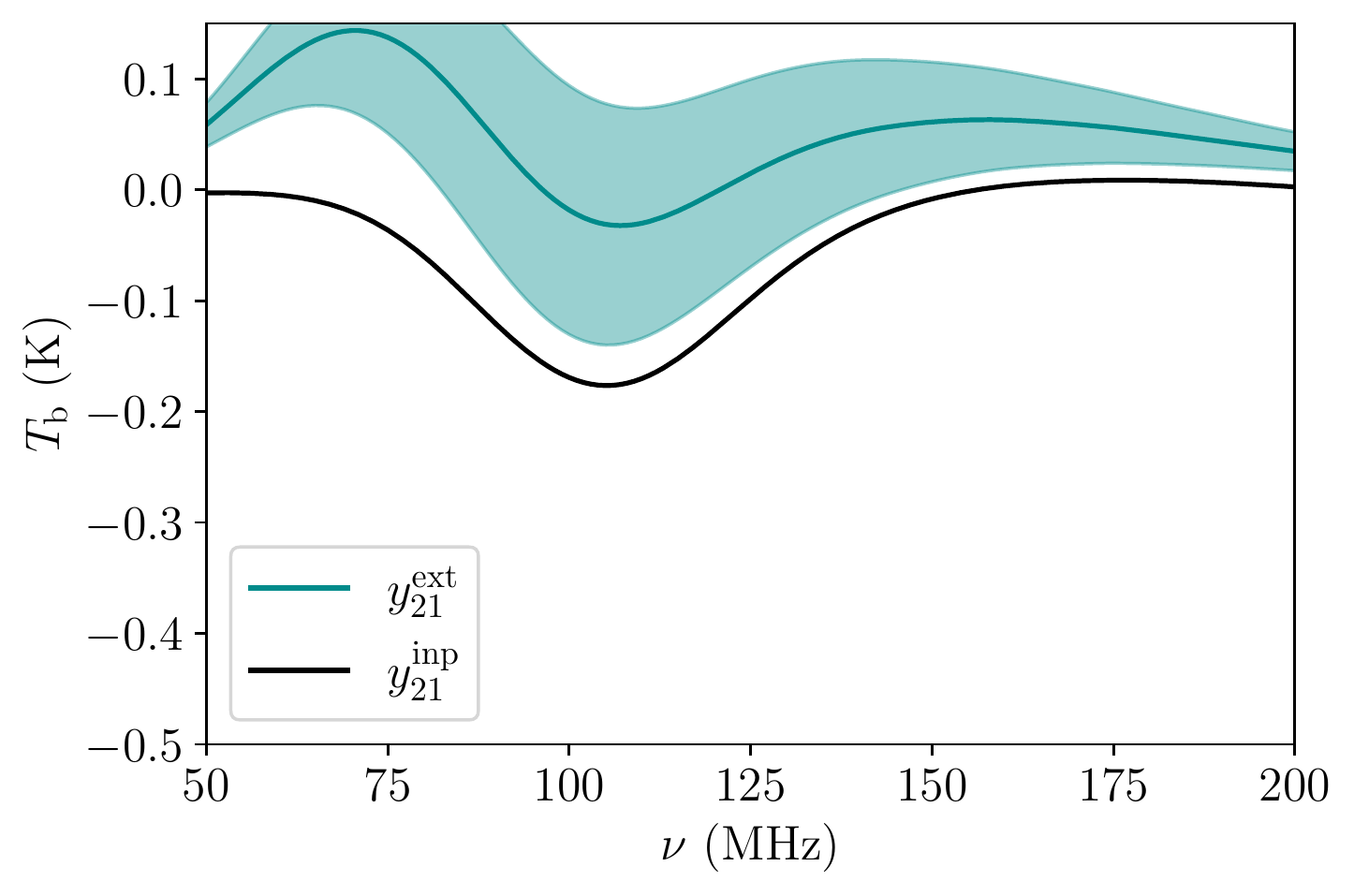}
    \includegraphics[width=0.32\linewidth]{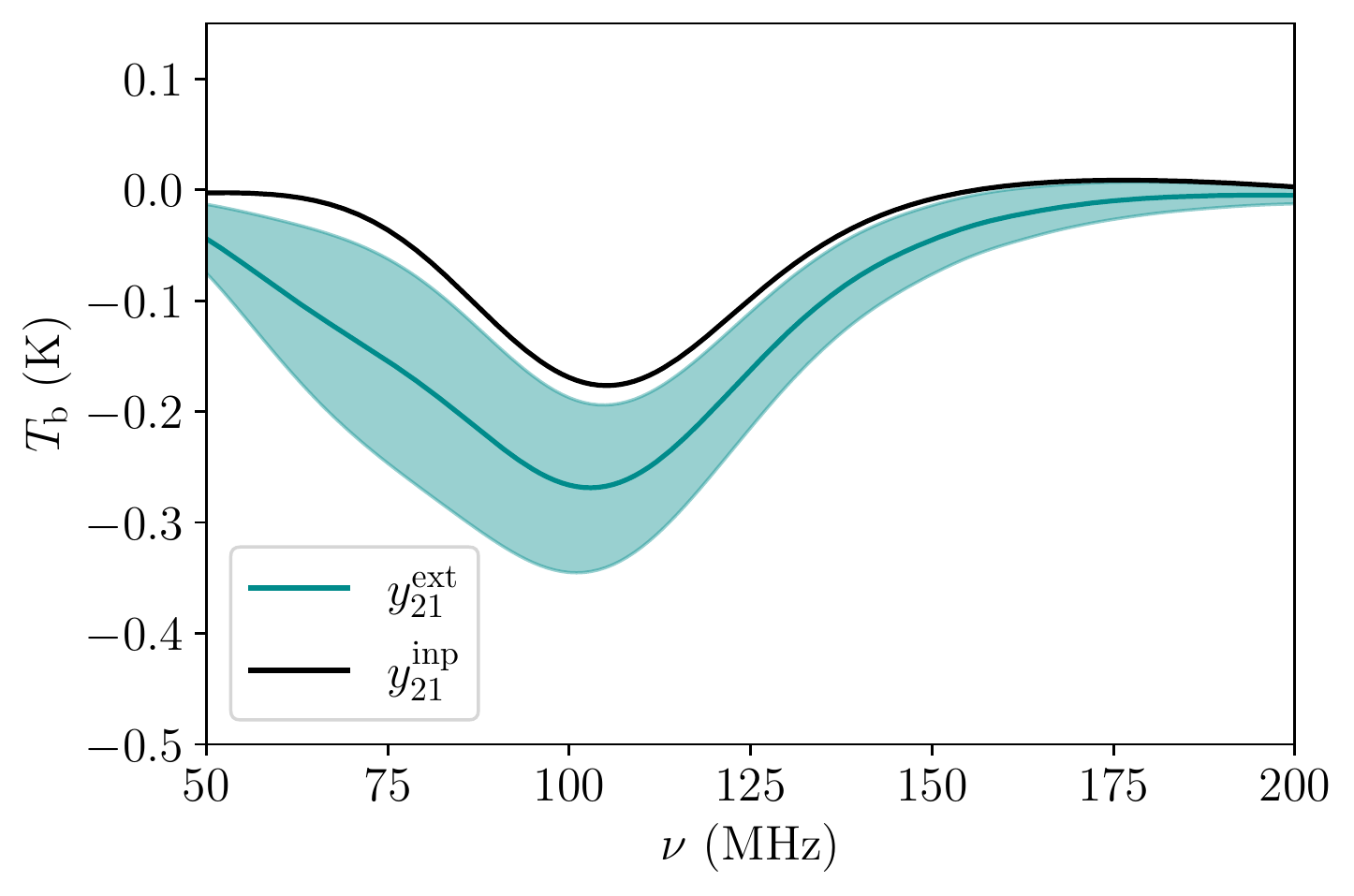}
    \includegraphics[width=0.32\linewidth]{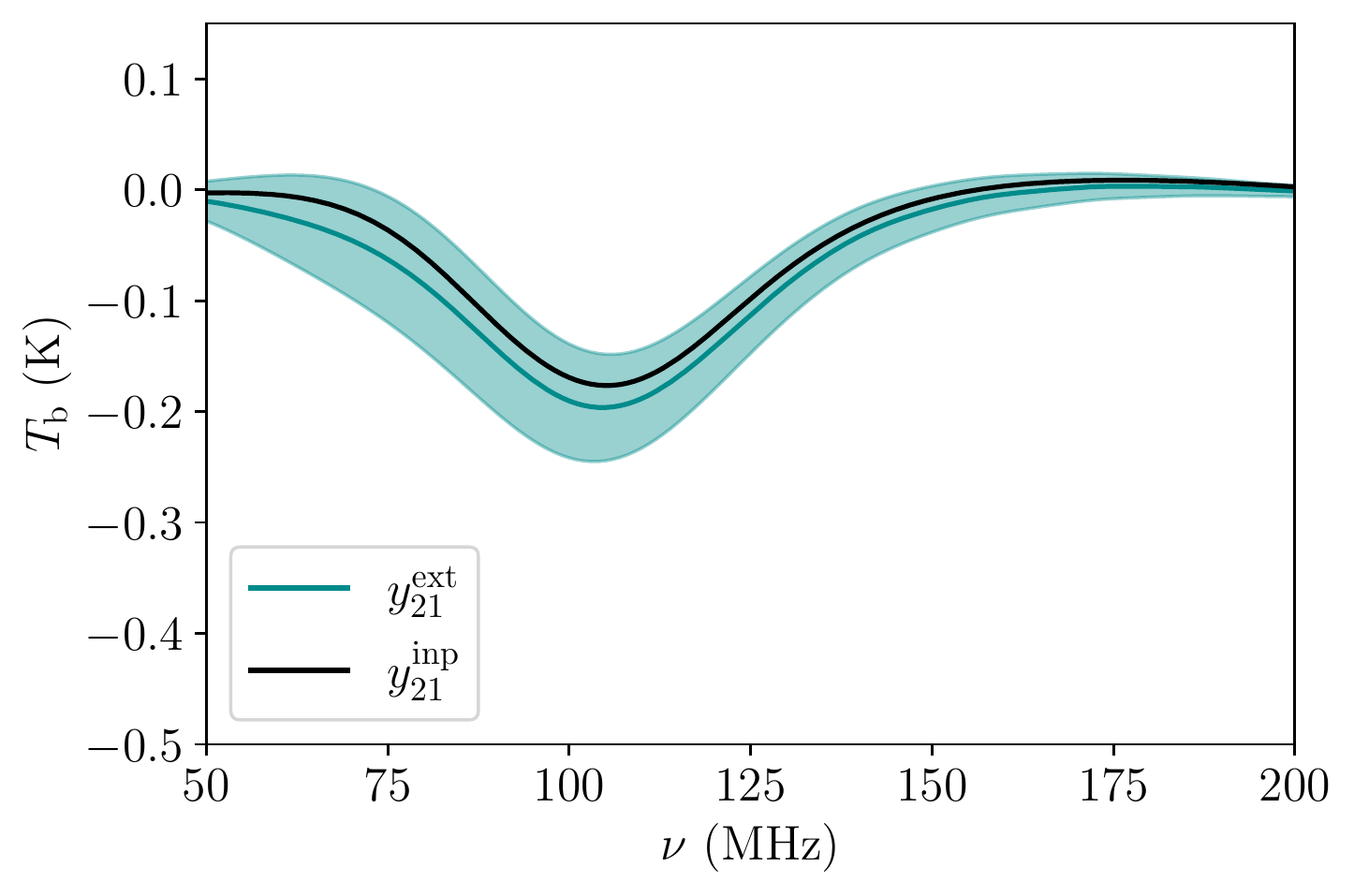}
    \caption{Extracted 21-cm signals for the mock observation generated from a physical 21-cm signal and a foreground model taken from the detailed foreground modeling set for dipole (left), dipole + conical log spiral (middle), and dipole + conical sinuous + conical log spiral (right) antennas configuration while simultaneously fitting four time bins.}
    \label{fig:ext_nreg_30}
\end{figure*}

To better understand the extent to which multiple antennas tend to improve the fit, we quantify the statistical measures of the extracted 21-cm signal that we discussed in Section~\ref{subsec:stat_meas}. First, we estimate the RMS uncertainty on the extracted signal for an ensemble of 5000 different mock observations, which are generated by randomly sampling a 21-cm and foreground component from their modeling sets. In the top panel of Figure~\ref{fig:stats_comp}, we show the cumulative distribution function (CDF) of the RMS uncertainty for seven different configurations that can be formed from the three antenna designs we considered. In each case, we fit the data from a single time slice and assume the total integration time to be 6 hours, which is distributed evenly in the analysis of multiple antennas. 
The advantage of having multiple antennas can be clearly noticed as the distribution for multiple antennas tends towards lower RMS uncertainty levels. For a single antenna, 68\% (and 95\%) of the 5000 different mock observations are extracted within 954 mK < RMS < 1808 mK (and 5568 mK < RMS < 8102 mK). For a pair of antennas, 68\% (and 95\%) of the mock observations are extracted within 55 mK < RMS < 71 mK (and 159 mK < RMS < 213 mK). For three antennas, 68\% (and 95\%) of the mock observations are extracted with RMS < 46 mK (and RMS < 126 mK). These values are tabulated in Table~\ref{tab:tab_simple} for different antenna configurations.
\addtolength{\tabcolsep}{-1pt}
\begin{table}
 \centering
 \caption{This tabulates the RMS uncertainty level (in mK) on the extracted 21-cm signal for 68\% and 95\% of the mock observations, and the variance of the PDF of $\chi^2$ for different antenna configurations shown in Figure~\ref{fig:stats_comp} for the simple foreground model.}
 \label{tab:tab_simple}
 \begin{tabular}{cccc}
  \hline
  Configuration & RMS|$_{68\%}$ & RMS|$_{95\%}$ & $\sigma$($\chi^2$)\\
  \hline
  \hline
  Dipole & 1807.64 & 8101.69 & 0.119\\
  Logspiral & 572.70 & 3054.16 & 0.118\\
  Sinuous & 954.46 & 5568.83 & 0.122\\
  \hline
  Dipole + Logspiral & 64.61 & 161.84 & 0.083\\
  Dipole + Sinuous & 55.96 & 159.41 & 0.083\\
  Logspiral + Sinuous & 70.52 & 212.96 & 0.085\\
  \hline
  Dipole + Logspiral + Sinuous & 45.83 & 125.49 & 0.067\\
  \hline
 \end{tabular}
\end{table}

In the bottom panel of Figure~\ref{fig:stats_comp}, we show the probability distribution function (from 5000 different mock observations) of the normalized $\chi^2$, which quantifies how well the set of foreground and 21-cm modes fit the data for all seven antenna configurations. As we include more antennas in the fit, the quality of the fit to the data improves as the variance of the $\chi^2$ distribution decreases, and it becomes more centered around $\chi^2=1$. For a single antenna, the variance of the distribution $\sigma(\chi^2) \sim 0.12$. For a pair of antennas, $\sigma(\chi^2) \sim 0.083$, which reduces to $\sigma(\chi^2) \sim 0.067$ for three antennas. These values for different configurations are tabulated in Table~\ref{tab:tab_simple}.

\subsection{Detailed Foreground Model}
Next, we discuss the results for a more realistic foreground model (see Section~\ref{subsec:detailed_modset}). In Figure~\ref{fig:ext_nreg_30}, we show the extracted 21-cm signal for three different antenna configurations: (i) dipole (left), (ii) dipole + conical log spiral (middle), and (iii) dipole + conical sinuous + conical log spiral (right). In each case, we simultaneously fit the data from four time bins, and the mock observation is generated by taking a 21-cm signal and a foreground component from their modeling sets.
\begin{figure}
    \centering
    \includegraphics[width=0.97\linewidth]{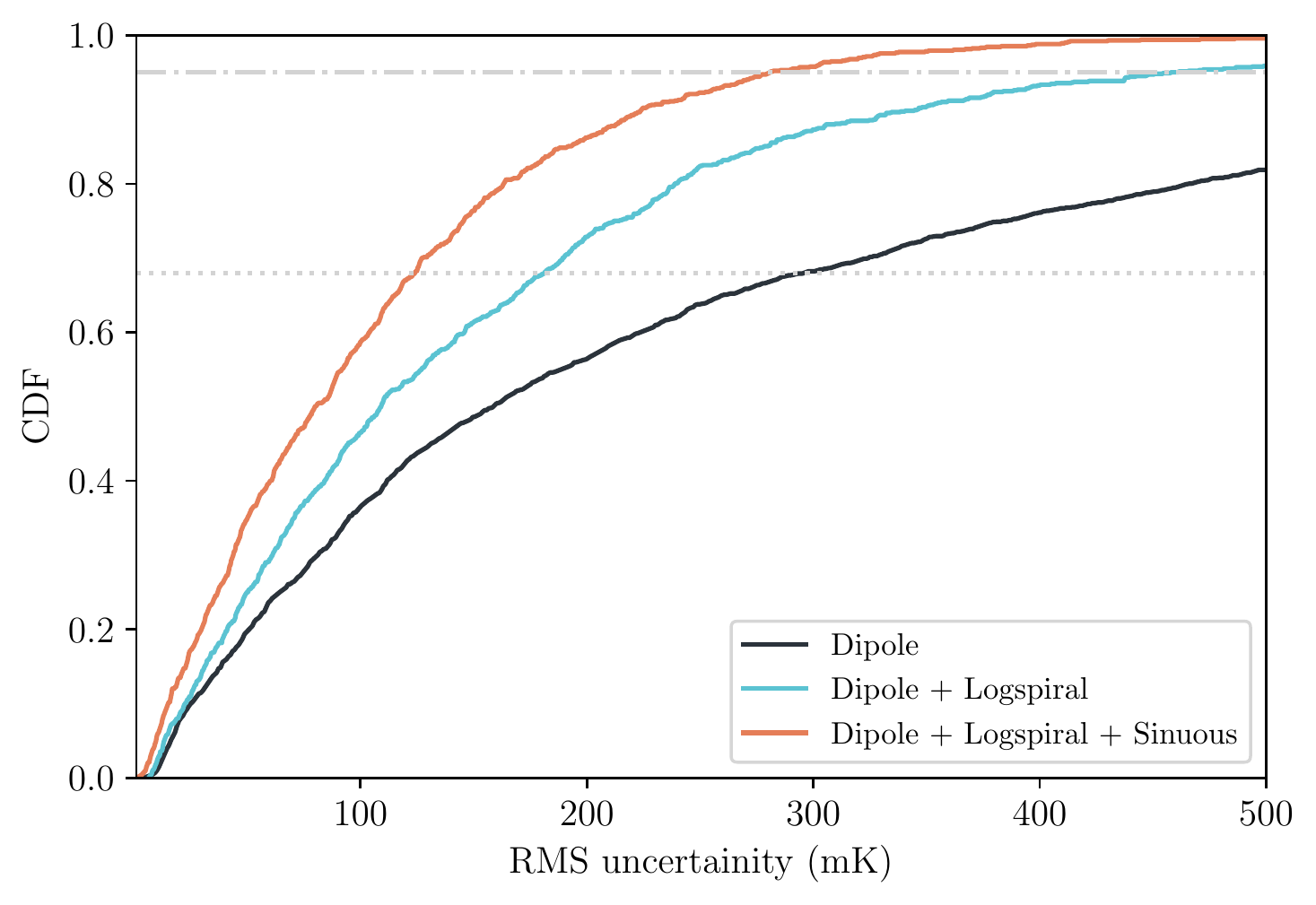}\vspace{0.4em}
    \includegraphics[width=0.97\linewidth]{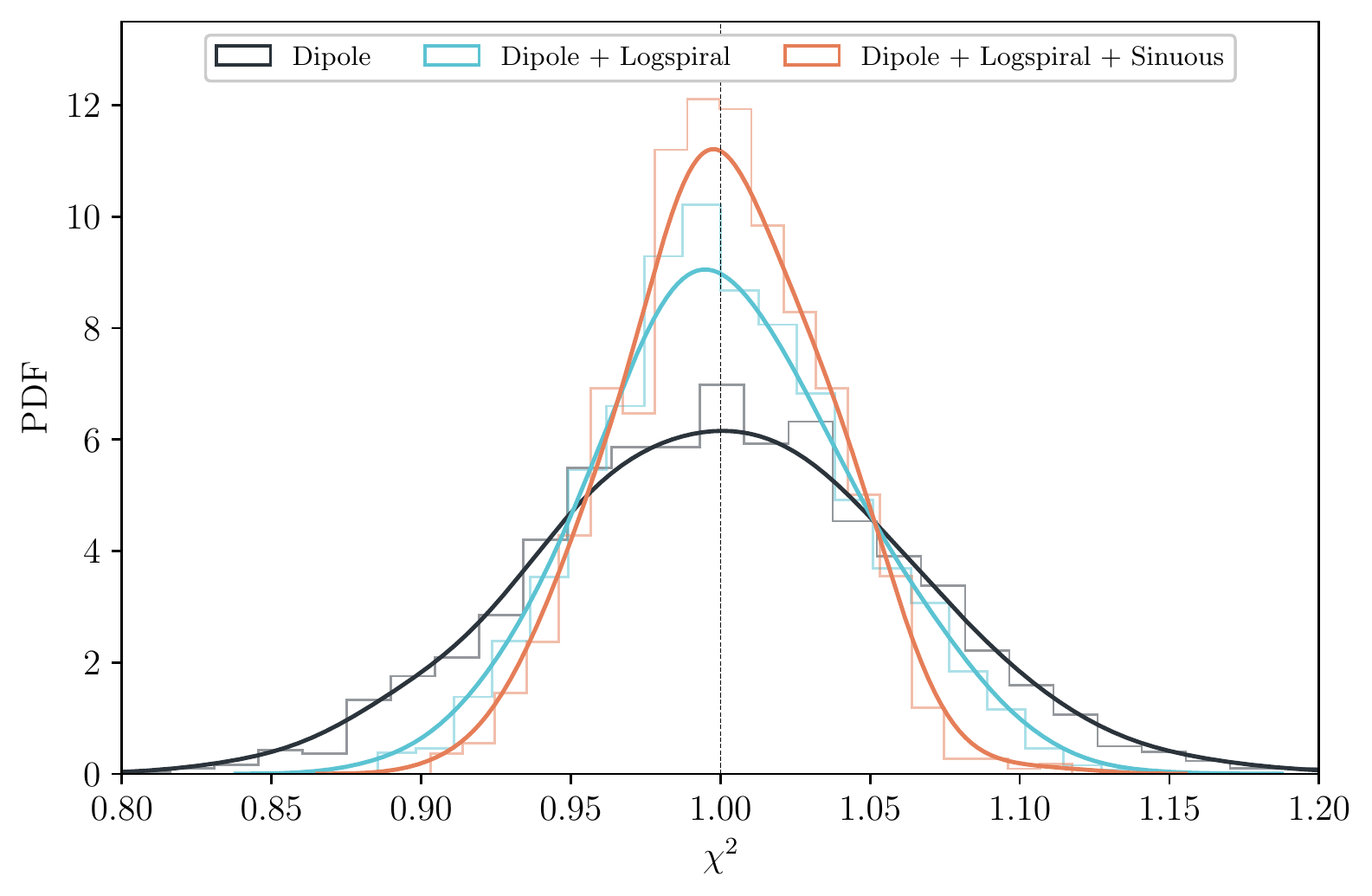}
    \caption{\textbf{Top panel:} Cumulative Distribution Function (CDF) of RMS uncertainty level on the extracted signal estimated from 2000 different mock observations for the detailed foreground model. Different colors represent different antenna configurations used in the analysis. The uncertainty level at 68\% and 95\% level are tabulated in Table~\ref{tab:detail_tab}. \textbf{Bottom panel:} Probability Distribution Function (PDF) of the normalized $\chi^2$ statistic estimated from 2000 different mock observations for each configuration.}
    \label{fig:stats_Nreg30}
\end{figure}

As expected, when considering a more realistic foreground model, the signal extraction is worse than it was for the simple foreground model, particularly for single antenna analysis. For example, if we compare the extracted signal from a dipole for the detailed foreground model (left panel of Figure~\ref{fig:ext_nreg_30}) with the extracted signal for the simple foreground model (bottom left panel of Figure~\ref{fig:ext_nreg_1}), the RMS uncertainty on the extracted signal is larger (123 mK in comparison to 21 mK). In addition, the extracted signal is also significantly biased. This happens because of a larger covariance between the foreground and the 21-cm signal modes, where the realistic foregrounds introduce an additional frequency-dependent structure in the higher-order foreground modes (see the bottom panel of Figure~\ref{fig:dFg_modset_basis}). Evidently, it is sub-optimal to use just a single antenna for 21-cm signal extraction with such foreground models.

We can improve this extraction if we utilize different antenna designs. For example, when we simultaneously fit the observations from a dipole and a conical log spiral antenna, as shown in the middle panel of Figure~\ref{fig:ext_nreg_30}, the uncertainty on the extracted signal decreases from 123 mK to 69 mK, and it also removes some of the bias in the extracted signal. This can be further improved by including all three antennas in the fitting, as shown in the right panel of the figure, where the RMS uncertainty on the extracted signal is 37 mK. 

We also estimate the CDF of the RMS uncertainty on the extracted signal from 2000 different mock observations formed from the detailed foreground model in the top panel of Figure~\ref{fig:stats_Nreg30} for three different antenna configurations. The distribution for multiple antennas tends towards lower RMS uncertainty levels. For three antennas, 68\% (and 95\%) of the mock observations are extracted with RMS < 124 mK (and < 281 mK) in comparison to RMS < 297 mK (and < 971 mK) for a single antenna. In the bottom panel, we show the PDF of the normalized $\chi^2$ statistic of the fit for the detailed foreground model. The variance of the $\chi^2$ distribution decreases from $\sigma(\chi^2) = 0.061$ to $0.034$ as we include more antennas in our fit. These values are tabulated in Table~\ref{tab:detail_tab}. These results suggest that utilizing different antenna designs is quite powerful for extracting cosmological signals when considering more complicated foreground models.

\begin{figure}
    \centering
    \includegraphics[width=\linewidth]{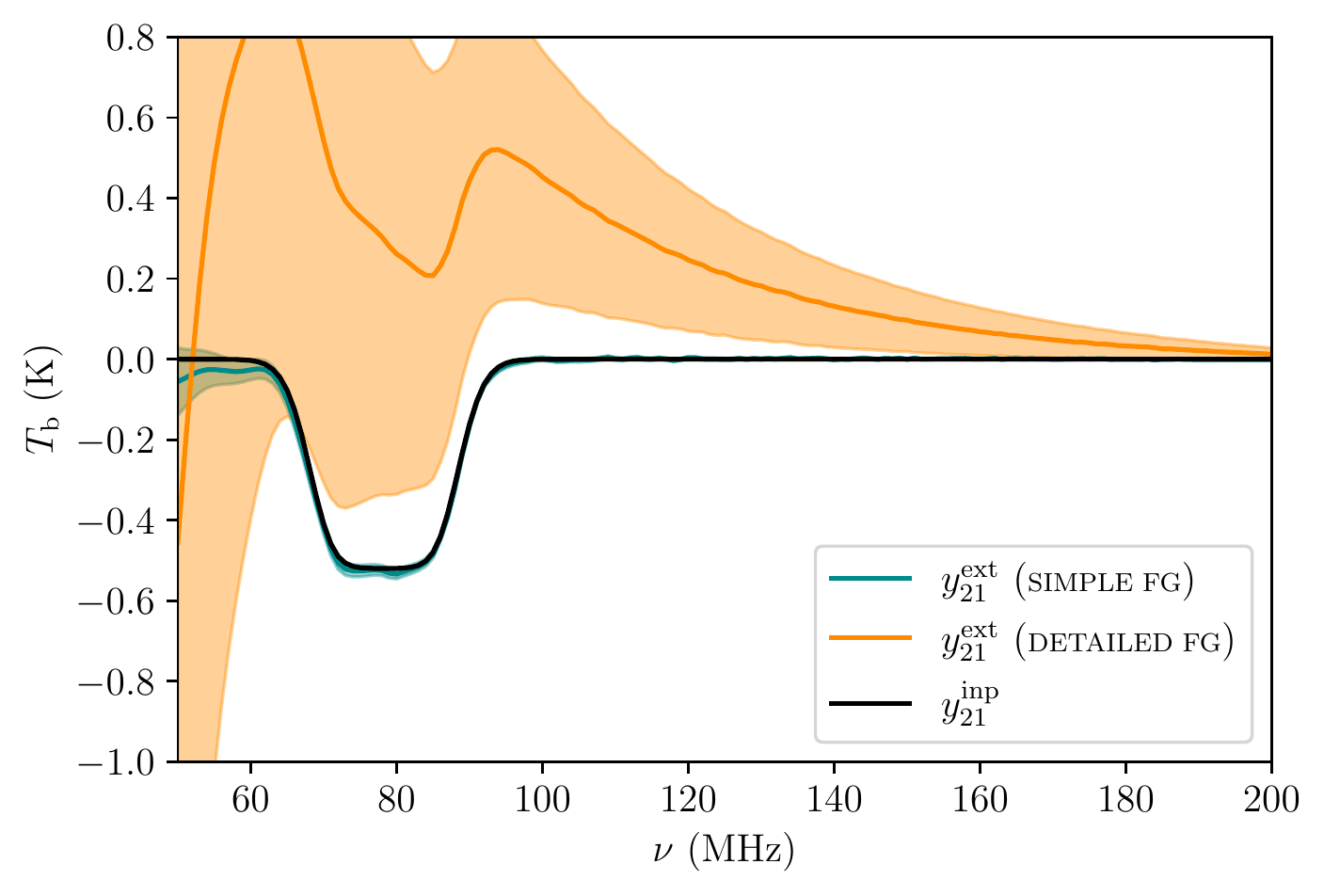}
    \caption{Extracted 21-cm signal for the simple (cyan) and detailed (orange) foreground model. The input 21-cm signal (black line) follows the EDGES profile and is not a part of the 21-cm modeling set.}
    \label{fig:outsideTs}
\end{figure}

\begin{table}
 \centering
 \caption{This tabulates the RMS uncertainty level (in mK) on the extracted 21-cm signal for 68\% and 95\% of the mock observations, and the variance of the PDF of $\chi^2$ for different antenna configurations shown in Figure~\ref{fig:stats_Nreg30} for the detailed foreground model.}
 \label{tab:detail_tab}
 \begin{tabular}{cccc}
  \hline
  Configuration & RMS|$_{68\%}$ & RMS|$_{95\%}$ & $\sigma$($\chi^2$)\\
  \hline
  \hline
  Dipole & 296.07 & 970.55 & 0.061\\
  \hline
  Dipole + Logspiral & 191.24 & 464.83 & 0.042\\
  \hline
  Dipole + Logspiral + Sinuous & 124.03 & 280.64 & 0.034\\
  \hline
 \end{tabular}
\end{table}
However, the performance of the signal extraction depends on how the 21-cm signal and foregrounds are simulated in their modeling sets. For example, in Figure~\ref{fig:outsideTs}, we show the extraction of the 21-cm signal when we take an EDGES like profile to render the mock observation while the 21-cm modeling set is simulated with only the physical signals. We simultaneously fit the data for all three antennas from four time slices. In this scenario, we can recover the 21-cm signal with the simple foreground model (cyan) with reasonable accuracy. The minimization of DIC, in this case, requires more 21-cm modes to fit the data. However, the signal extraction is significantly biased for our detailed foreground model (orange) because of the more significant covariance between the foreground and 21-cm modes. This could become substantially more challenging when one includes the uncertainties in the foreground model as well.

\section{Summary}
\label{sec:summary}
In this article, we have shown a method to utilize multiple different antenna designs in the global 21-cm signal extraction. Our analysis is based on the current approach of the REACH experiment, but it can easily be generalized to any global 21-cm signal experiment. Our formalism for extracting the 21-cm signal is based on fitting the modes derived from the Singular Value Decomposition of the 21-cm signal, and beam-weighted foreground modeling sets \citep{Tauscher_2018,Tauscher_2020_util}.

We demonstrated the impact of utilizing multiple antennas to better constrain the 21-cm signal by considering two different foreground models (i) A simple foreground model, where we assumed the spectral index to be constant across the sky and used that spectral index to create a modeling set and (ii) A more realistic foreground model, where we considered a variation of the spectral index across the sky, divided into 30 regions.

We find that for the simple foreground model, we can extract the signal with any antenna design when we simultaneously fit the data for multiple time slices. This extraction is further improved when we include multiple different antennas in our fitting. To better understand this improvement, we quantify some statistical measures of the extracted 21-cm signal for an ensemble of different mock observations. For multiple antennas, the cumulative distribution function of the RMS uncertainties tends towards lower RMS uncertainty levels. In addition, the variance of the probability distribution function decreases as we include more antennas, and it becomes more centered around $\chi^2=1$.

The impact of including multiple antennas in the fitting becomes more pronounced when we consider the more realistic foreground model in the mock observation. In that case, with only a single antenna, even when we fit the data for multiple time bins, we can not precisely recover the 21-cm signal. Both the large RMS of the reconstructed signal and the significant bias show that only using a single antenna is likely, not sufficient to recover the cosmological signal. Instead, when we use multiple antennas of different types, we reduce the bias in the extracted signal and also the RMS uncertainty level by a factor of 2-3.

The signal extraction presented in this analysis could be further improved by measuring all four Stokes parameters from different antennas instead of only Stokes $I$ \citep{Tauscher_2020_util}. This could further reduce the covariance between the signal and foreground modes and could be crucial for analyzing more detailed foreground models, which overlap more significantly with the 21-cm models. This will be considered in a future work. We also plan to utilize the spectral constraints derived in this paper to calculate the full posterior probability distribution of any signal parameter space of choice while marginalizing over the linear foreground modes \citep{Rapetti_2020}, which results in a tighter constraint on the 21-cm signal. To derive these constraints, one can quickly scan the parameter space by using the sky-averaged 21-cm signal emulators \citep{Cohen_2020, Bye_2022}.
 
\section*{Acknowledgements}
We thank the Center for Information Technology of the University of Groningen for their support and for providing access to the Peregrine high performance computing cluster. P.D.M acknowledges support from the Netherlands organization for scientific research (NWO) VIDI grant (dossier 639.042.730). LVEK acknowledges the financial support from the European Research Council (ERC) under the European Union’s Horizon 2020 research and innovation programme (Grant agreement No. 884760, "CoDEX").

\section*{Data Availability}
Accompanying code is available at \url{https://github.com/anchal-009/SAVED21cm}. The data underlying this article will be shared on reasonable request to the corresponding author.

\bibliographystyle{mnras}
\bibliography{refer}



\appendix

\bsp	
\label{lastpage}
\end{document}